\documentclass[11pt]{article}
\usepackage{amssymb,amsmath,mathrsfs,enumerate}
\usepackage{graphicx,rotate,multicol}
\usepackage{float}
\usepackage{tocloft}
\usepackage{subfig}
\usepackage[margin=10pt,labelfont=bf]{caption}
\usepackage{cite}
\usepackage{soul}
\usepackage{slashed}
\usepackage{multirow}
\usepackage[colorlinks=true,
linkcolor=blue,
urlcolor=blue,
citecolor=red]{hyperref}

\makeatletter
\let\Hy@linktoc\Hy@linktoc@page
\makeatother
\evensidemargin=\oddsidemargin
\newcommand{\lsim}{\mbox{\raisebox{-.6ex}{~$\stackrel{<}{\sim}$~}}}
\newcommand{\gsim}{\mbox{\raisebox{-.6ex}{~$\stackrel{>}{\sim}$~}}}

\allowdisplaybreaks

\title{\bf Searching for Leptoquarks at IceCube and the LHC}

\author{\bf Ujjal Kumar Dey,$^{a,}$\footnote{ujjal.dey1@gmail.com} 
	\hspace{4pt} Deepak Kar,$^{b,}$\footnote{deepak.kar@cern.ch} 
	\hspace{4pt} Manimala Mitra,$^{c,d}$\footnote{manimala@iopb.res.in} \\
	 {\bf Michael Spannowsky,}$^{e}$\footnote{michael.spannowsky@durham.ac.uk}
	\hspace{4pt} {\bf Aaron C. Vincent}$^{f,g}$\footnote{aaron.vincent@queensu.ca}
	\\[10pt]
	\small\em $^a$Centre for Theoretical Studies, Indian Institute of Technology Kharagpur,\\
	\small\em Kharagpur 721302, India\\
	\small\em $^b$School of Physics, University of the Witwatersrand,\\
	\small\em Johannesburg, Wits 2050, South Africa\\
	\small\em $^c$Institute of Physics, \\
	\small\em Sachivalaya Marg, Bhubaneswar, Odisha 751005, India\\
	\small\em $^d$Homi Bhabha National Institute, \\
	\small\em Training School Complex, Anushakti Nagar, Mumbai 400085, 
	          India\\
	\small\em $^e$Institute for Particle Physics Phenomenology, Durham University, \\
	\small\em Durham DH1 3LE, United Kingdom\\
	\small\em $^f$Department of Physics, Imperial College London,
Blackett Laboratory, \\
    \small\em Prince Consort Road SW7 2AZ, United Kingdom\\
	\small\em $^g$Canadian Particle Astrophysics Research Centre (CPARC),\\
	\small\em Department of Physics, Engineering Physics and Astronomy, \\
	\small\em Queen’s University, Kingston, Ontario K7L 3N6, Canada 		
		}

\date{}

\begin{document}

%\preprint{IP/BBSR/2017-11\\
%IPPP/17/47}

\maketitle

\begin{abstract}
In the light of recent experimental results from IceCube, LHC searches for scalar leptoquark, and the flavor anomalies $R_K$ and $R_{K^*}$, we analyze  two scalar leptoquark models with hypercharge $Y=1/6$ and $Y=7/6$. We consider the 53 high-energy starting events from IceCube and perform a statistical analysis, taking into account both the Standard Model and leptoquark contribution together. The lighter leptoquark states that are in agreement with IceCube are strongly constrained from LHC di-lepton+dijet search. Heavier leptoquarks in the TeV mass range are in agreement both with IceCube and LHC. We furthermore show that leptoquark which explains the $B$-physics anomalies and does not have any coupling with the third generation of quarks and leptons, can be strongly constrained.
\end{abstract}

\newpage

\hrule \hrule
\tableofcontents
\vskip 10pt
\hrule \hrule

%%%%%%%%%%%%%%%%%%%%%%%%%%%%%%%%%%%%%%%%%%%%
\section{Introduction \label{intro}}
%%%%%%%%%%%%%%%%%%%%%%%%%%%%%%%%%%%%%%%%%%%%
The recent discovery of the Higgs boson \cite{Aad:2012tfa,Chatrchyan:2012xdj} completes the Standard Model (SM)  of particle physics as an effective, well-tested, theory up to the energy range around the electroweak scale. However, it fails to explain apparent observations in Nature, e.g. the observed neutrino mass,  matter-antimatter asymmetry or dark matter, and it cannot provide a fundamental reason for the quantisation of the particles' charges or a possible unification of the standard model gauge groups. 
Thus, the primary task of experiments that can access the high-energy regime is to search for new particles and interactions which are necessary ingredients to the extensions of the SM.

Leptoquarks (LQ), particles that simultaneously carry lepton number (L) and baryon number (B), are predicted by a large variety of new physics scenarios, in particular by grand unified theories (GUTs) \cite{Georgi:1974sy,Fritzsch:1974nn}, the Pati-Salam model \cite{Pati:1974yy} and extended technicolor models \cite{Farhi:1980xs,Gabriel:1993ai}. Assuming that leptoquark interactions with SM particles are $SU(3)_C \times SU(2)_L \times U(1)_Y$ invariant and have dimensionless couplings, there are only twelve different types of leptoquarks, according to the possible assignments of their respective quantum numbers \cite{Buchmuller:1986zs, Buchmuller:1986iq, Dorsner:2016wpm}. Half of the leptoquark types are of scalar nature and the other half of vector nature. Yet, in each case the coupling structure to different SM generations can be highly complex, unless constrained by the UV theory from which the leptoquark descends. 

Historically, experimental searches for leptquarks have been a high priority and have been performed at a wide range of collider experiments, e.g. at HERA \cite{Aid:1995wd,Abramowicz:2012tg}, the Tevatron \cite{GrossoPilcher:1998qt}, LEP \cite{Abreu:1993bc} or the LHC \cite{Aad:2015caa, Aaboud:2016qeg, Khachatryan:2016jqo,Sirunyan:2017yrk}. The ATLAS and CMS collaborations searched for leptoquarks coupled predominantly to the first two generation SM fermions, focusing on final states with two electrons or muons and two jets \cite{Aad:2015caa, Aaboud:2016qeg, CMS:2014qpa, CMS:2016qhm, CMS:2016imw}. Absence of deviations from SM backgrounds limits the leptoquark mass to $M_{LQ} \geq 1100$ GeV at 90$\%$ C.L. \cite{Aaboud:2016qeg}. Additionally, searches by CMS have further constrained third-generation leptoquarks to $M_{LQ} \geq 850$ GeV  \cite{Sirunyan:2017yrk, Khachatryan:2016jqo}. Previous bounds on their masses were $M_{LQ} > 699$ GeV from HERA \cite{Abramowicz:2012tg} and $M_{LQ} > 225$ GeV from the Tevatron \cite{GrossoPilcher:1998qt}. 

Interestingly, recent flavor anomalies, indicating the violation of lepton-flavor universality in rare $b$-transitions measured by ATLAS \cite{ATLAS-CONF-2017-023}, CMS \cite{CMS-PAS-BPH-15-008} and LHCb \cite{Aaij:2014pli, Aaij:2014ora, Aaij:2015esa, Aaij:2015yra, Aaij:2017vbb, Das:2016vkr, Das:2017kfo}, have reignited the interest in leptoquark phenomenology \cite{Sakaki:2013bfa, Hiller:2014yaa, Gripaios:2014tna, Varzielas:2015iva, Becirevic:2015asa, Fajfer:2015ycq, Becirevic:2016yqi, Aaij:2015oid, Wehle:2016yoi, Becirevic:2016oho, Das:2016vkr, Crivellin:2017zlb, Becirevic:2017jtw, Hiller:2017bzc, Cai:2017wry, Aloni:2017ixa,DiLuzio:2017vat}.

{Other than the collider constraints, leptoquarks can also be searched  in the high-energy neutrino-neucleon interactions, namely  at the IceCube Neutrino Observatory in Antarctica  \cite{Aartsen:2013jdh,Aartsen:2014gkd,Aartsen:2015zva}. The four year high-energy starting events (HESE) sample \cite{Aartsen:2014gkd, Aartsen:2015zva},  reported by the IceCube collaboration span an energy range from  22 TeV to 2 PeV. The three events above a PeV in energy have raised a considerable  amount of interests in  recent years \cite{Barger:2013pla, Bhattacharya:2014yha, Dutta:2015dka, Dey:2015eaa, Dev:2016uxj, Dev:2016qbd, Dey:2016sht, Mileo:2016zeo, Bhattacharya:2016tma, Dev:2016qeb, Kistler:2016ask, Chauhan:2017ndd, Bhattacharya:2017jaw, Borah:2017xgm, Palladino:2017qda}. A primary  goal of this work is to constrain leptoquark mass and coupling using all of the HESE sample. We also critically reexamine the idea that scalar leptoquarks can provide a natural explanation to the three events seen above a PeV  at the IceCube.} Indeed, the centre of mass energy scale of a PeV neutrino colliding with a nucleon at rest is just the right scale to excite a $\sim$ TeV leptoquark resonance \cite{Anchordoqui:2006wc}.  Many  studies have made the claim that such a process can improve the fit to the observed event rate  \cite{Barger:2013pla, Dutta:2015dka, Dey:2015eaa, Dev:2016uxj, Dey:2016sht, Mileo:2016zeo}. These are often framed in terms of an excess that requires a non-standard explanation; however, it should be stressed that the observed event rate falls within the expected Waxman-Bahcall flux \cite{1999PhRvD..59b3002W}, consistent with a common origin with cosmic rays (see also \cite{Waxman:2015ues} for a review). 
 
{In this work, analysing the leptoquark contribution to the IceCube events,  we pinpoint few  major improvements as compared to the existing analysis.} %that greatly reduce the sensitivity to such a signal. 
These include 1) the inclusion of the whole energy range 10 TeV -- 10 PeV, including the energy bins in which no events have been observed; and 2) a simultaneous fit of new physics and SM contribution to the 53 observed events in 4 years of data. While point 1) generally reduces the goodness of fit of any new component on top of the SM event rate, point 2) represents a very crucial improvement: we will find that by allowing the spectral index and normalizations of the incoming astrophysical flux and atmospheric backgrounds to vary freely, it is not possible to  single out a LQ signal while remaining consistent with the model still allowed by LHC data. Any small improvement in the goodness of fit furthermore remains much lower than the 1$\sigma$ level. 

This paper is structured as follows: we begin by describing the details of the two benchmark models that we consider in Sec. \ref{sec:models}. We then examine the effect of these models on the observed neutrino event rate at IceCube in Sec. \ref{icecube1}. We present the exclusions from LHC and the discussion on flavor anomalies in Secs. \ref{sec:LHC} and \ref{sec:flavor}, respectively. Appendices \ref{app:modelA} and \ref{app:modelB} provide the detailed cross sections used for our IceCube analysis. We conclude in Sec. \ref{sec:conclusion}.

%%%%%%%%%%%%%%%%%%%%%%%%%%%%%%%%%%%%%%%%%%%%
\section{Models} \label{sec:models}
%%%%%%%%%%%%%%%%%%%%%%%%%%%%%%%%%%%%%%%%%%%%
In this section we briefly review the models of interest: A) scalar leptoquark $\chi_1$ having the representations $(\mathbf{3}, \mathbf{2}, \mathbf{1/6})$, and B) scalar leptoquark $\chi_2$ having the representation $(\mathbf{3}, \mathbf{2}, \mathbf{7/6})$  under the SM gauge group $SU(3)_{c}\otimes SU(2)_{L} \otimes U(1)_{Y}$.  $\chi_1$  can be represented as $\chi_1=(\chi^{2/3}_1, \chi^{-1/3}_1)^{\rm T}$, where $\chi^{2/3}$ and $\chi^{-1/3}$ are the components of $SU(2)_{L}$ doublets; the superscripts represent their corresponding electromagnetic charges. The other leptoquark $\chi_2$ has the form $\chi_2=(\chi^{5/3}_2, \chi^{2/3}_2)^{\rm T}$. Below, we discuss the Lagrangian for the two leptoquarks mentioned above: 

\begin{itemize}
\item
The leptoquark $\chi_1 (\mathbf{3}, \mathbf{2}, \mathbf{1/6})$: The Yukawa interaction of $\chi_1$ with SM fermions can be given as, 
\begin{align}
\mathcal{L}_{\Phi} = 
              &- x_{ij} \bar{d}^i_{R} {\chi_1}. \ell_{L}^{j} 
               + {\rm h.c.} \notag \\
            = &- x_{ij} \bar{d}^i P_{L} l^{j} \chi^{2/3}_1
                + x_{ij} \bar{d}^i P_{L} \nu^{j} \chi^{-1/3}_1
                + {\rm h.c.},
\label{eq:yukawar21a}
\end{align}
where $\ell_{L}$ is the $SU(2)_{L}$ doublet lepton, $l=(e, \mu, \tau)$ are the charged leptons,  and $i,j~(=1,2,3)$ represent the generation indices of the leptons. In addition to
the Yukawa Lagrangian, the field $\chi_1$ has the kinetic and mass terms 
\begin{equation}
\mathcal{L}_k=({D^{\mu}{\chi_1}})^{\dagger}({D_{\mu}\chi_1})~~~~ \mathrm{and}~~~~\\
\mathcal{L}_m=\frac{1}{2} m^2_{\chi_1}\chi^{\dagger}_1\chi_1\\
\label{kinmassa1}
\end{equation}
respectively, where $D_{\mu}=\partial_{\mu}-ig_sT^a G^a-igW^i_{\mu}\frac{\sigma_i}{2}-i\frac{g^{\prime}}{12}$ denotes the covariant derivative, and $m_{\chi_1}$ is the mass of the leptoquark. 
Note that the leptoquark $\chi_1$ interacts only with down type of quarks ($d, s, b$). 

\item
The leptoquark $\chi_2 (\mathbf{3}, \mathbf{2}, \mathbf{7/6})$: 
The leptoquark $\chi_2$  has similar $SU(3)_c$ and $SU(2)_L$ charges, and non-trivial hypercharge $\mathbf{7/6}$. The Lagrangian of this leptoquark field has the following form:
\begin{align}
\mathcal{L}_{\Phi} = 
              &- x_{ij} \bar{u}^i_{R} {\chi_2}. \ell_{L}^{j} + y_{ij} \bar{e}^i_R\chi^*_2Q^j_L
               + {\rm h.c.} \notag \\
            = &- x_{ij} \bar{u}^i P_{L} l^{j} \chi^{5/3}_2
                + x_{ij} \bar{u}^i P_{L} \nu^{j} \chi^{2/3}_2+(yV^{\dagger})_{kl}\bar{e}^k P_L u^l\chi^{-5/3}_2+y_{kl} \bar{e}^k P_L d^l\chi^{-2/3}_2
                + {\rm h.c.}.
\label{eq:yukawa76a}
\end{align}
In the above, the quarks and charged leptons are written in their mass basis and $V$ is the CKM matrix.
The kinetic and mass terms of leptoquark are 
\begin{equation}
\mathcal{L}_k=({D^{\mu}{\chi_2}})^{\dagger}({D_{\mu}\chi_2}), ~~~~\\
\mathcal{L}_m=\frac{1}{2} m^2_{\chi_2}\chi^{\dagger}_2\chi_2,\\
\label{kinmassa2}
\end{equation}
where $D_{\mu}$ is the covariant derivative $D_{\mu}=\partial_{\mu}-ig_sT^a G^a-igW^i_{\mu}\frac{\sigma_i}{2}-i\frac{7 g^{\prime}}{12}.$

\end{itemize}

The leptoquarks $\chi_1$ and $\chi_2$ have interactions with the SM fermions. As we will discuss in the next sections, the leptoquark components $\chi^{1/3}_1$ and $\chi^{2/3}_2$ can have observable consequences at IceCube. 

\begin{figure}[t]
\begin{center}
\includegraphics[scale=0.5]{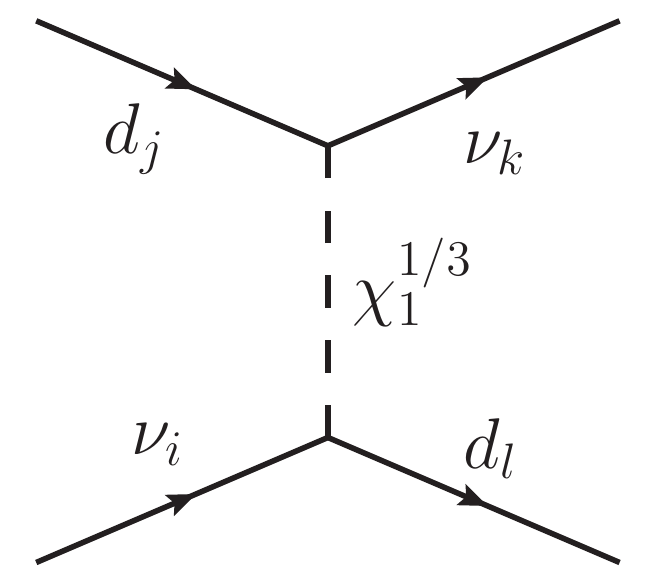}~~~~~~~~
\includegraphics[scale=0.5]{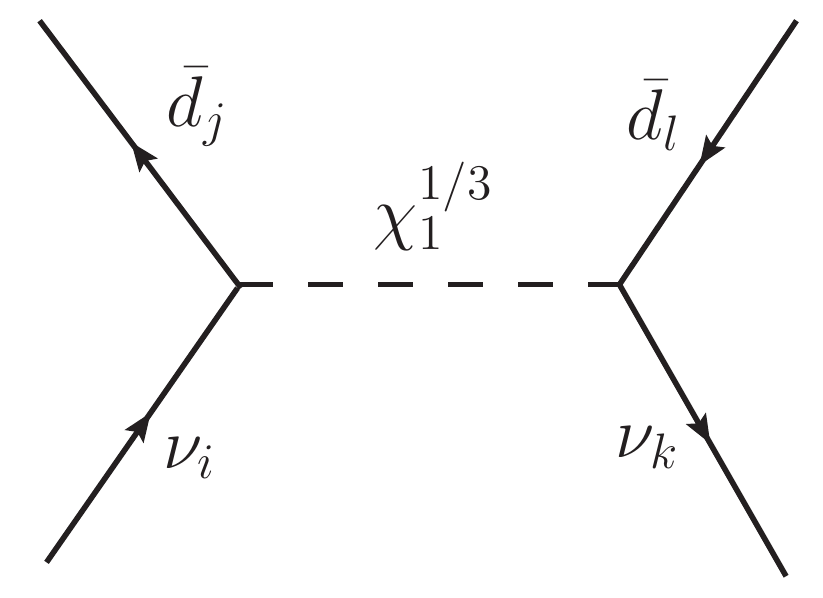}
\caption{Feynman diagrams for neutrino-quark interactions through LQ $\chi^{1/3}$, relevant for Model A.}
\label{f:feyndiag1}
\end{center}
\end{figure}

\begin{figure}[t]
\begin{center}
\includegraphics[scale=0.4]{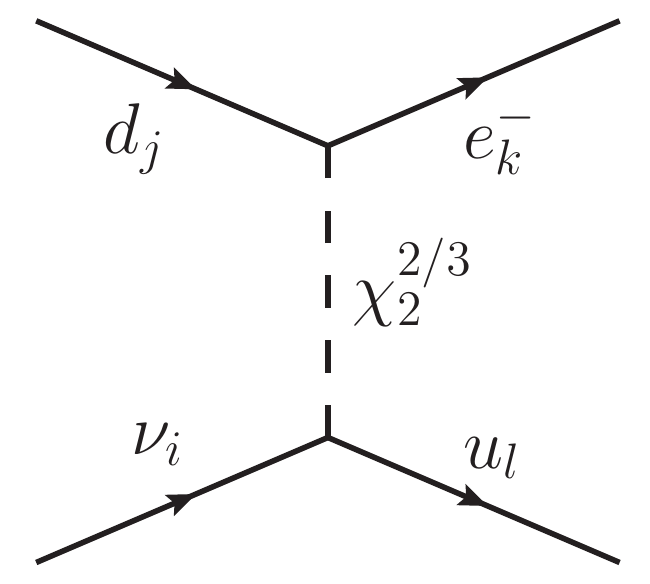}~~~
\includegraphics[scale=0.4]{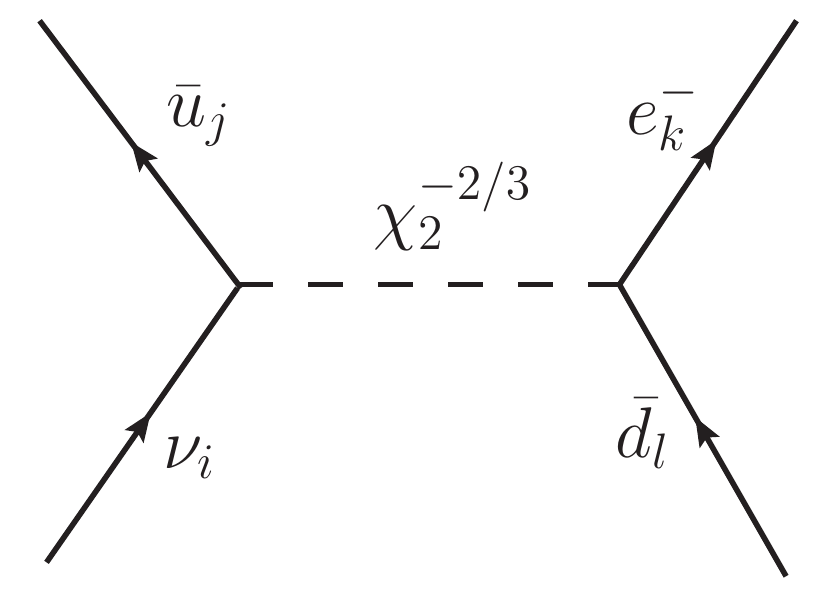}~~~
\includegraphics[scale=0.4]{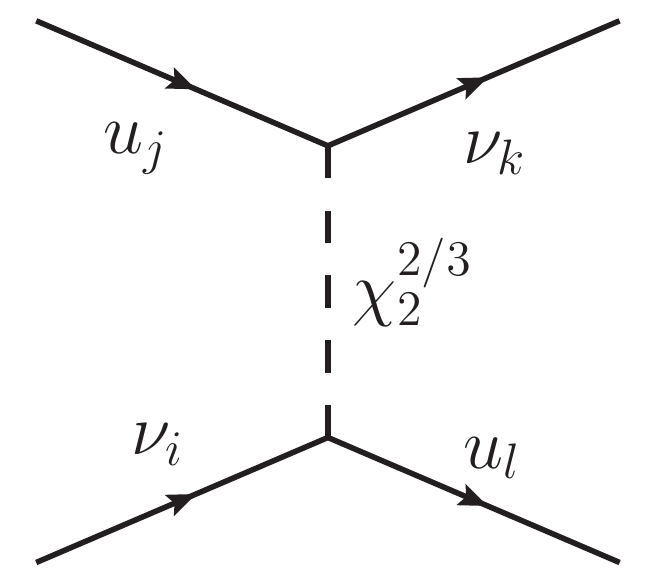}~~~
\includegraphics[scale=0.4]{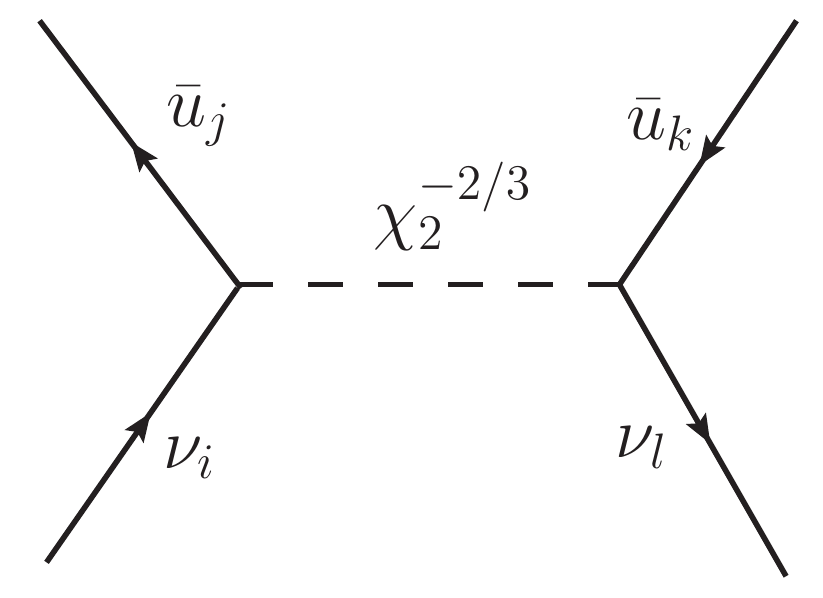}
\caption{Feynman diagrams for neutrino-quark interactions through LQ $\chi^{2/3}$, relevant for Model B.}
\label{f:feyndiag2}
\end{center}
\end{figure}

%%%%%%%%%%%%%%%%%%%%%%%%%%%%%%%%%%%%%%%%%%%%
\section{IceCube events \label{icecube1}}
%%%%%%%%%%%%%%%%%%%%%%%%%%%%%%%%%%%%%%%%%%%%

In this section we discuss the interaction of neutrinos (and antineutrinos) with detector protons and neutrons, mediated by the heavy
leptoquarks, with the goal of improving the spectral fit to the observed high-energy PeV IceCube events.  The leptoquark $\chi^{-1/3}_1$ mediates the neutral current interactions $\nu_{i}d_{j}\to \nu_{k}d_{l}$ and $\nu_{i}\bar{d}_{j}\to \nu_{k}\bar{d}_{l}$. We show the Feynman diagram for these two processes in Fig.~\ref{f:feyndiag1}. The leptoquark $\chi^{2/3}_2$ has both the neutral current as well as  charged current interactions, $\nu_i u \to \nu_i u$, $\nu_i \bar{u} \to e d$, $\nu d \to e u$ as shown in Fig.~\ref{f:feyndiag2}. Note that, the leptoquark $\chi^{5/3}_2$ does not interact with the light neutrinos, and hence does not contribute to the IceCube neutrino-nucleon interaction. For Model A, the relevant processes are 
\begin{itemize}
\item 
 $\bar{\nu } {d} \to \bar{\nu} {d}$ mediated by s-channel LQ, and similarly $\nu \bar{d} \to \nu \bar{d}$,
\item
$\nu d \to \nu d$ mediated by the t-channel  LQ, and similarly $\bar{\nu} \bar{d} \to \bar{\nu} \bar{d}$. 
\end{itemize}
In addition to the above, the SM contribution has both the  $W^{\pm}$ mediated charged current (CC),   and $Z$ mediated neutral current (NC) contributions that give rise to $\nu d \to \nu u $, $\nu u \to \nu d/ l$, $\nu d \to \nu l$, $\nu u \to \nu l$ and $\nu d/u \to \nu d/u$ interactions. The expressions for neutrino-nucleon cross sections are given for both NC and CC processes in Appendices \ref{app:modelA} and \ref{app:modelB}.
\begin{figure}
\includegraphics[scale=0.3]{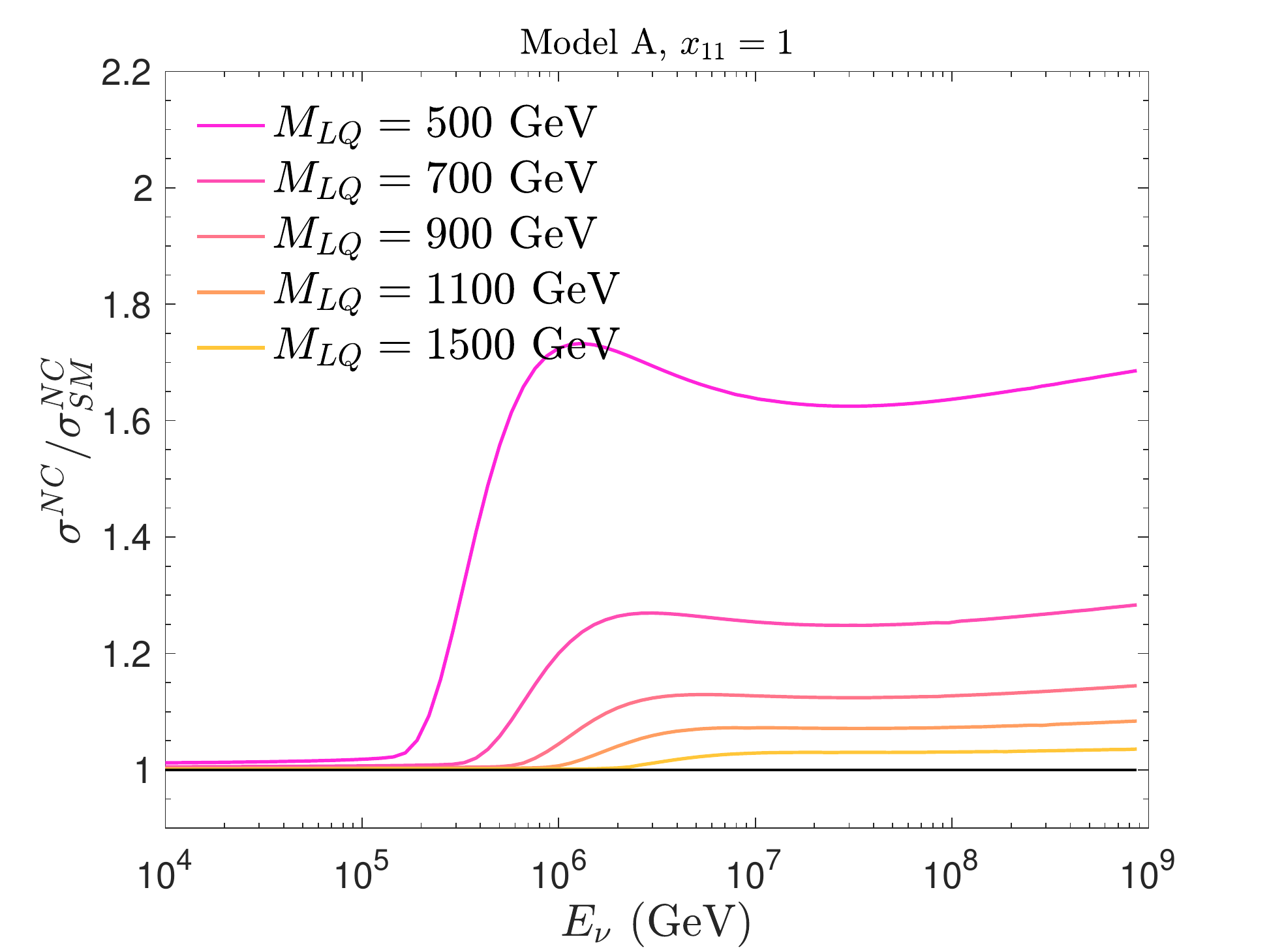}~
\includegraphics[scale=0.3]{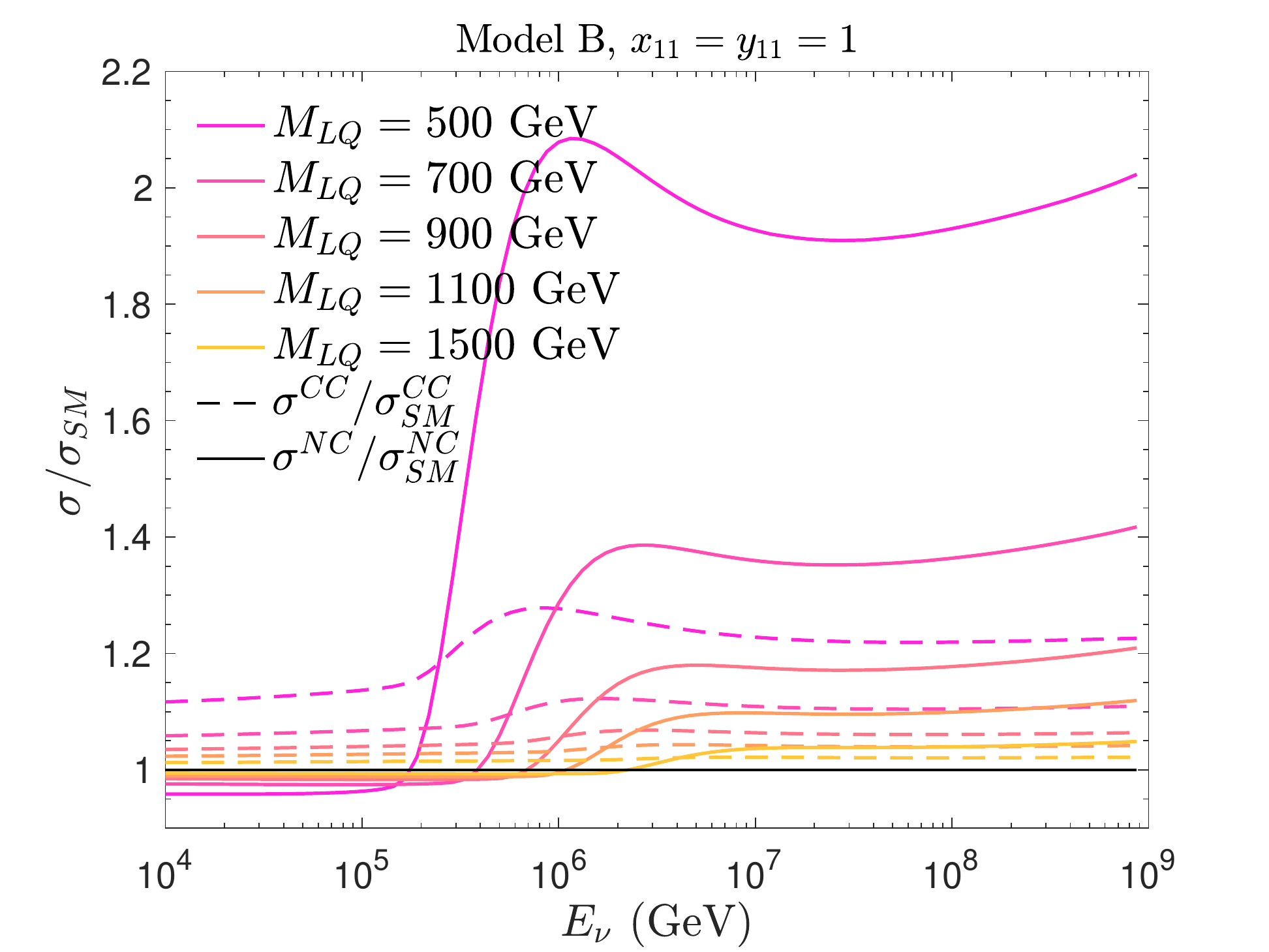}~
\includegraphics[scale=0.3]{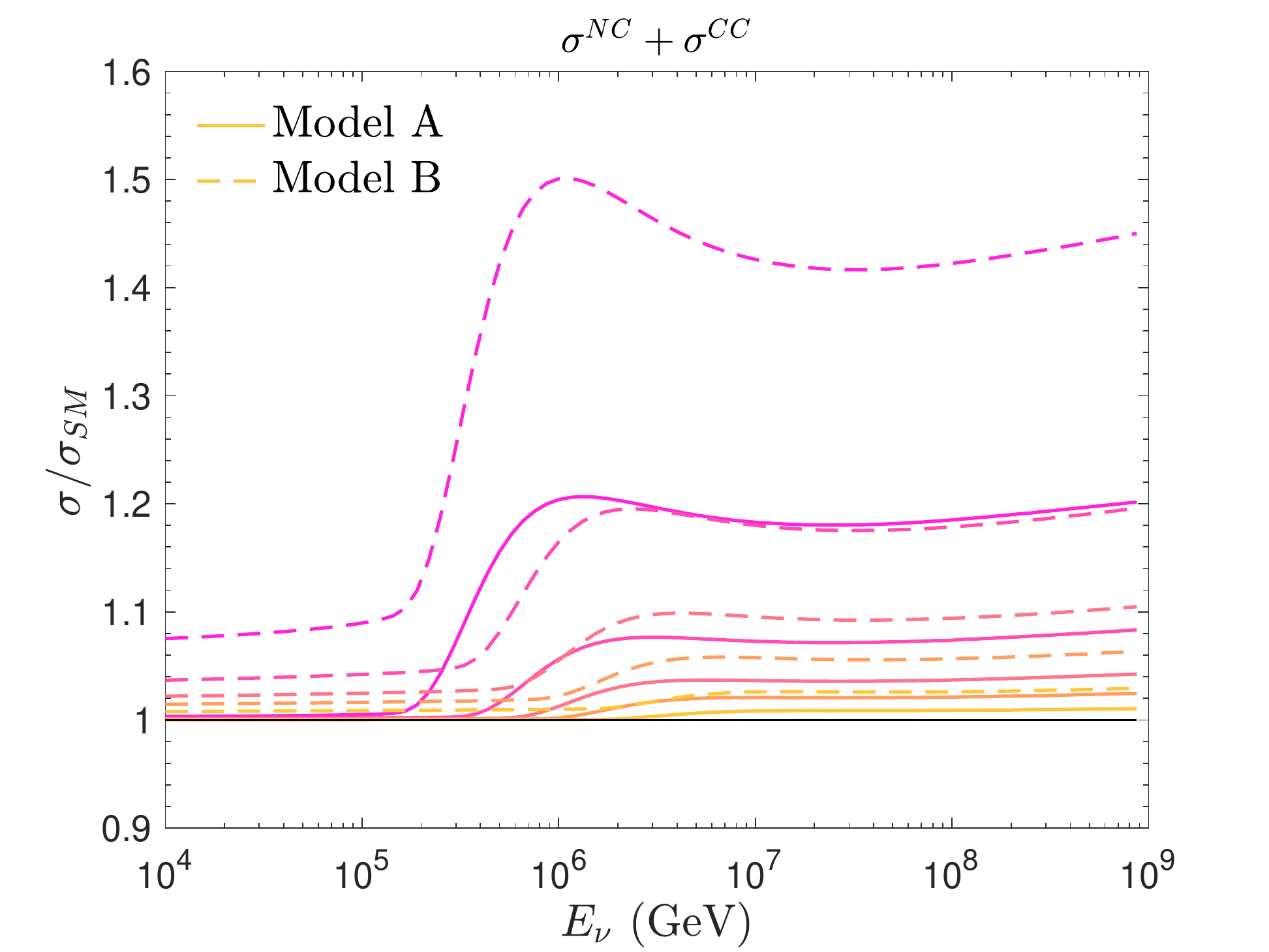}
\caption{The ratio of neutrino-nucleon cross sections, between the Standard Model (SM) + leptoquark model and the relevant Standard Model  only channel, for five different LQ masses. Left: Model A, neutral current contribution only ($x_{11}= 1$, $x_{12} = x_{13}  = 0.1$). Middle: Model B neutral current contribution only (solid); charged current contribution only (dashed) ($x_{11} = y_{11} = 1$, {all other couplings set to zero}). Right: ratio of the total cross section to the SM case (note different $y$ scale). }
\label{fig:LQxs}
\end{figure}
The leptoquark $\chi^{-1/3}_1$ can only produce shower events, while the leptoquark $\chi_{2}^{2/3}$ can produce both shower and muon track events if $y_{kl}$ is nonzero for $k = 2$ or $3$. For this analysis, we only allow the $k = 1$ final-state charged lepton (i.e. no final-state muons or taus), so that both CC and NC interactions lead only to showers. 

\subsection{Energy deposition rates at IceCube}
We calculate the number of events at IceCube  following the formalism  laid out in \cite{Palomares-Ruiz:2015mka}. The relevant observable at IceCube is the event rate per deposited EM-equivalent energy. We distinguish between the incoming neutrino energy $E_\nu$, the ``true'' EM-equivalent energy $E_{true}$ deposited in the ice, and the deposited energy \textit{in the detector,} $E_{dep}$. The latter two are related by a Gaussian smearing function $R(E_{true},E_{dep}, \sigma)$, with a width $\sigma(E_{true}) = 0.112E_{true}$. The neutral current event rate for an incoming neutrino of flavour $j$ is:
\begin{align}
\frac{d N^{sh,NC}_{\nu_j}}{dE_{dep}} = T N_A \int_0^\infty  Att_{\nu_i}(E_\nu) \frac{d\phi_{\nu_j}(E_\nu)}{dE_\nu} \int_0^1 dy M_{eff}(E_{true}) \nonumber \\
\times ~R(E_{true},E_{dep},\sigma(E_{true})) \frac{d \sigma^{NC}_{\nu_j}(E_\nu,y)}{dy}.
\label{eq:dNdEshowernue}
\end{align}
In the above, $d\phi_{\nu_j}/dE_\nu$ is the astrophysical neutrino flux; the flavour dependence of the cross section  $\sigma_{\nu_j}$ arises from the leptoquark couplings $x_{1j}$. $T$ is the exposure time (1347 days for the four-year HESE sample), and $N_A$ is Avogadro's number, representing the number of target nucleons per gram of ice\footnote{Remembering that ice is not isoscalar; rather: $n:p = 4:5$ must be used when computing cross sections.}. At energies above 10 TeV, the neutrino mean free path becomes smaller than the diameter of the Earth, which is included in the attenuation factor $Att_{\nu_i}(E_\nu)$. To properly take this into account, one should compute the directionally-dependent attenuation rate by solving the transport equation. We will use angle-averaged attenuation rates computed by \cite{Palomares-Ruiz:2015mka}; Ref.~\cite{Mileo:2016zeo} has shown that the impact of LQs on attenuation is quite small.  $M_{eff}(E_{true})$ is the effective detector mass as a function of the true electromagnetic equivalent energy, and can be seen as the fiducial IceCube volume times an energy-dependent detector efficiency. A parametrization of this function can be found in Ref. \cite{Palomares-Ruiz:2015mka}. Though IceCube provides effective masses as a function of the incoming neutrino energies, these depend implicitly on the $\nu-N$ cross sections: since we are searching for effects of new physics, the method of Eq. (\ref{eq:dNdEshowernue}) must be used. Finally, the true energy $E_{true} = E_h$, where the hadronic energy $E_h = F_h(E_\nu y)E_\nu y$. For $F_h$ we use the parametrization from Ref \cite{Gabriel:1993ai}:
\begin{equation}
F_h(E_X) = 1 - (1-f_0)\left(\frac{E_X}{E_0}\right)^{-m},
\end{equation}
with $f_0 = 0.467$, $E_0 = 0.399$ GeV and $m = 0.130$ \cite{Kowalski}. 

For Model B, CC events can also occur. If the final state is an electron, then the deposition rate is the same as Eq. \eqref{eq:dNdEshowernue}, but with the cross section replaced with the relevant CC cross section (detailed in Appendix \ref{app:CC}), and with $E_{true} = E_h + E_\nu(1-y)$ to take into account the electron's energy deposition.

Finally, we include other Standard Model processes: tau neutrino CC interactions, which can produce both showers and muon tracks, as well as muon neutrino CC interactions, which produce tracks. We furthermore include electron antineutrino interactions with ice electrons, which become dominant around the Glashow resonance at $E_\nu = 6.3$ PeV~\cite{Glashow:1960zz}.  Though we are interested in new physics contributions to the total cascade rate, we include the Standard Model track events as a way of constraining the total astrophysical flux. The details of these rates are in Appendix B of \cite{Palomares-Ruiz:2015mka}. We employ the CT14QED \cite{Schmidt:2015zda} PDF sets, accessed via LHAPDF6 \cite{Buckley:2014ana}.

The dominant production mechanism for astrophysical neutrinos is expected to be charged pion decay, from high energy $pp$ and $p\gamma$ collisions. These produce neutrinos in a ratio $(\nu_e:\nu_\mu:\nu_\tau)$ = $(1:2:0)$. After oscillation over large, uncorrelated distances, this should average to a composition very close to $(1:1:1)$ which we take to be the flavour composition of our astrophysical neutrino flux. Known oscillation physics prevents large deviations from this flavour ratio, and, in fact, as long as production remains pion-dominated, it remains difficult to stray far from the $(1:1:1)$ composition in new physics scenarios \cite{Arguelles:2015dca}. 

In addition to astrophysical neutrinos, a non-negligible fraction of the HESE sample is composed of atmospheric neutrinos, as well as veto-passing muons from atmospheric showers.  We take the shape of the atmospheric shower spectrum from \cite{Vincent:2016nut}\footnote{We note that the LQ interactions should also affect the atmospheric neutrino detection rate; however, this is a subdominant effect for the LQ masses under consideration here, given the lower energy range of the atmospheric contribution ($\lesssim 60$ TeV). We therefore take the post-interaction atmospheric background from \cite{Vincent:2016nut}, thereby avoiding some of the systematics (e.g. self-veto rates) involved in modelling the atmospheric neutrino flux.}. The atmospheric muon flux uses the parametrization of Eq.~(1) from the same reference.

Finally, it has been shown \cite{Palomares-Ruiz:2015mka,Aartsen:2015ivb,Vincent:2016nut} that a non-zero misidentification rate of tracks as showers (e.g. from large muon inelasticity, events near the detector edge, etc.) affects reported event topologies. This was notably the reason behind the lower-than-expected track-to-shower ratio in the HESE data sets since the first two years' data \cite{Mena:2014sja,Palomares-Ruiz:2014zra,Vincent:2015woa}. We include this by making the replacements to expected track (tr) and shower (sh) event rates $\lambda$:
\begin{eqnarray}
\lambda^{tr} &\rightarrow& (1-p_{mID})\lambda^{tr}, \\
\lambda^{sh} &\rightarrow& \lambda^{sh}  + p_{mID}\lambda^{tr},
\end{eqnarray}
with the mis-ID probability is set to $p_{mID} = 0.2$. This is lower than the published value of 0.3 \cite{Aartsen:2015ivb}, which also included lower-energy (MESE) events for which track reconstruction is inherently more difficult. We note that a proper quantification of $p_{mID}$ has never been performed, and we will return to the effects of varying this quantity in Sec. \ref{sec:ICresults}.

\subsection{Comparison with IceCube data}
\label{sec:ICdata}
Out of the 53 events\footnote{Although 54 were reported, one was a coincident muon track whose energy could not be reconstructed.} in the 4-year HESE  sample, 14 are muon tracks and 39 are cascades. Among these roughly 14 are expected to be from atmospheric muons, and 7 from atmospheric neutrino events \cite{Aartsen:2014gkd,Aartsen:2015zva}. We are interested in the effect of a new scalar leptoquark on the goodness of fit of the model (SM+LQ) to the 4-year publicly available data. We make two crucial changes with respect to previous studies:
\begin{enumerate}
\item We simultaneously fit the background (Standard Model) contribution, by allowing the following four parameters to vary: the astrophysical spectral index $\gamma$, the astrophysical flux normalization $\phi_0$, the atmospheric neutrino rate $N_{\nu,atm.}$, and the veto-passing atmospheric muon rate $N_\mu$.
\item We include all energy bins from 10 TeV to 10 PeV, including bins in which zero events were observed. We separate this range into 15 logarithmically spaced bins for showers, and the same number for muon tracks. This distinction is important, as the different contributions to the event rates have very different topological signatures.
\end{enumerate}
 These are crucial:  previous studies \cite{Barger:2013pla, Dutta:2015dka, Dey:2015eaa, Dey:2016sht, Mileo:2016zeo} have examined the impact of an extra leptoquark-induced interaction (or a similar R-parity violating SUSY model \cite{Dev:2016uxj}) on the best-fit fluxes reported by the IceCube Collaboration. However, this is not self-consistent, as the fluxes were derived using the same data set. Indeed, there is a large degeneracy between the flux normalization, spectral component, and new physics contribution; these must all be taken into account simultaneously, along with the atmospheric flux normalizations. The inclusion of zero-event bins is also critical, as we will find that any improvement to the fit to the three observed PeV events is accompanied by a worse fit to those empty bins, especially at and above the 6.3 PeV Glashow resonance.
 
We also do not employ the narrow width approximation (NWA), as was done in previous studies, and instead use the full expressions (\ref{eq:dsignubar}, \ref{eq:dsignu3}), which include terms that mix SM and LQ contributions. We do note that for couplings $\lesssim 1$, the NWA does provide fairly accurate corrections to the neutrino-nucleus cross sections.

We take the event rate in each bin to follow Poisson statistics. For each point in $(M_{LQ},x_{ij})$ parameter space, we first maximise the fit with respect to the four nuisance parameters ($\gamma,\phi_0,N_{\nu,atm},N_{\mu}$), before evaluating the goodness of fit, parametrized via the \textit{p}-value. This is done via a Monte Carlo event simulation of the Poisson process. 

\subsection{Results}
\label{sec:ICresults}
Even with known model parameters, the systematic uncertainties in energy deposition make the enhanced cross sections shown in Fig.~\ref{fig:LQxs} difficult to distinguish from the standard case. An unknown inelasticity $y$ for each event means that only a fraction of the neutrino energy might end up deposited in the ice, which smears the detected spectrum -- in addition to the $\sim 10\%$ error on reconstructed event energy $E_{dep}(E_{true})$. 

We perform the fitting procedure detailed above, including event energy and topology information from the 53 IceCube HESE events. In the case where no new physics interactions are present (SM only), we find a best-fit spectrum 
\begin{equation}
\frac{d\phi_{astro}}{dE_\nu} = \phi_0 \left(\frac{E_\nu}{100 \mathrm{\, TeV}}\right)^{-\gamma} 
\end{equation}
with $\gamma = 2.8$ and $\phi_0 = 7.3 \times 10^{-18}$ GeV$^{-1}$ cm$^{-2}$ s$^{-1}$ sr$^{-1}$. Backgrounds are best fit by 10.7 atmospheric neutrinos, and 4.9 veto-passing atmospheric muons. 

\begin{figure}[t]
%\begin{tabular}{c c c}
\includegraphics[scale=0.3]{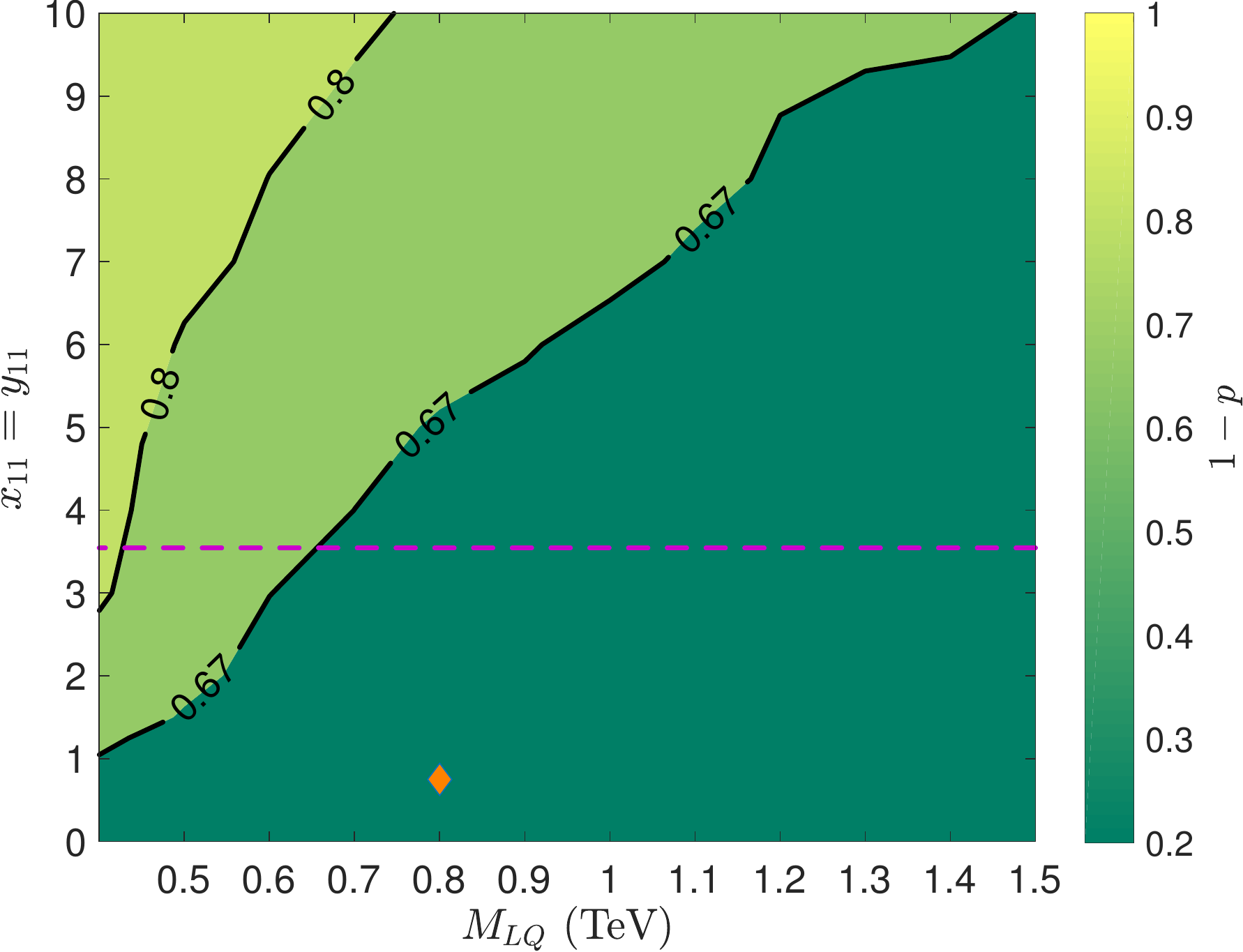} ~ %& 
\includegraphics[scale=0.3]{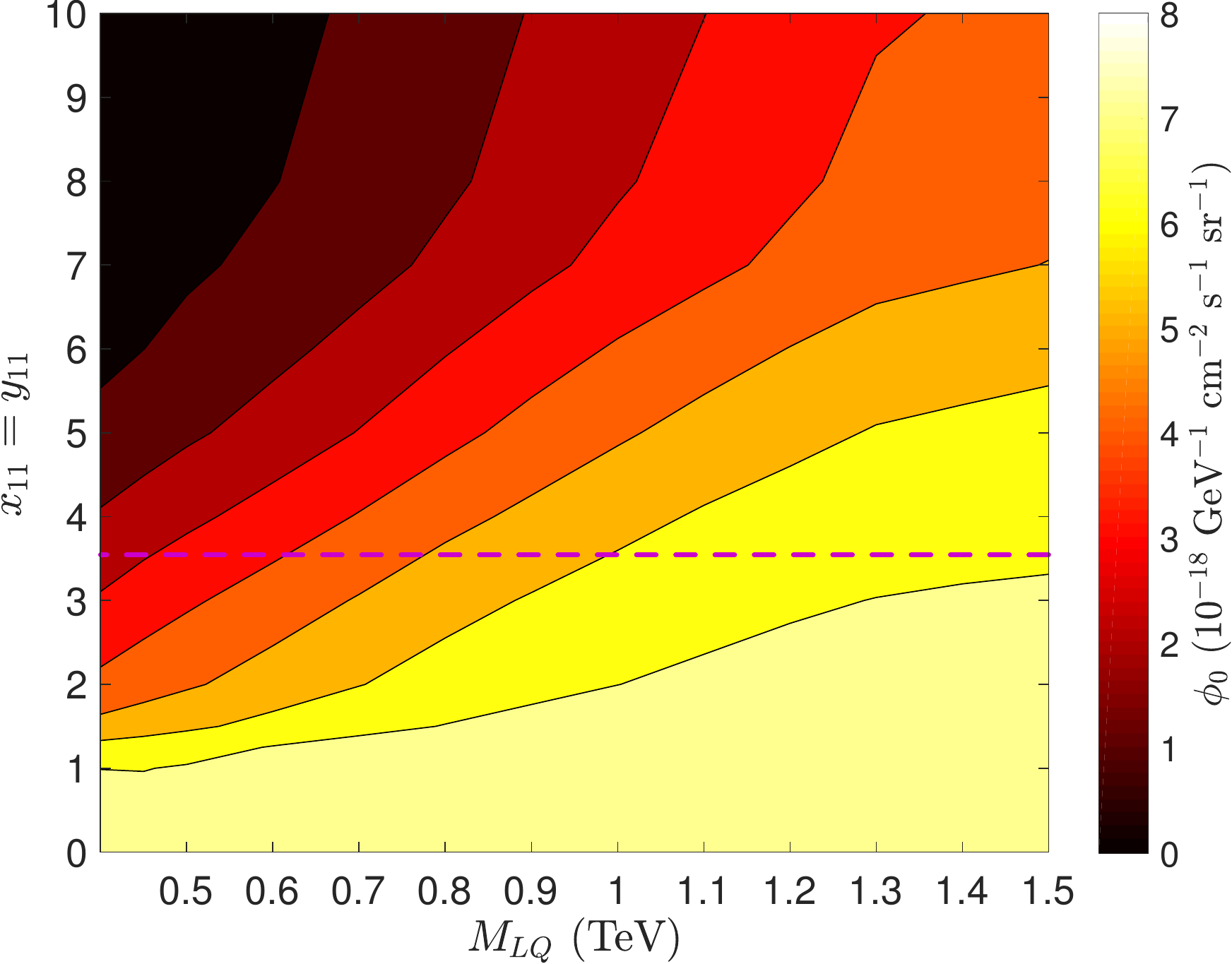} ~ %&
\includegraphics[scale=0.3]{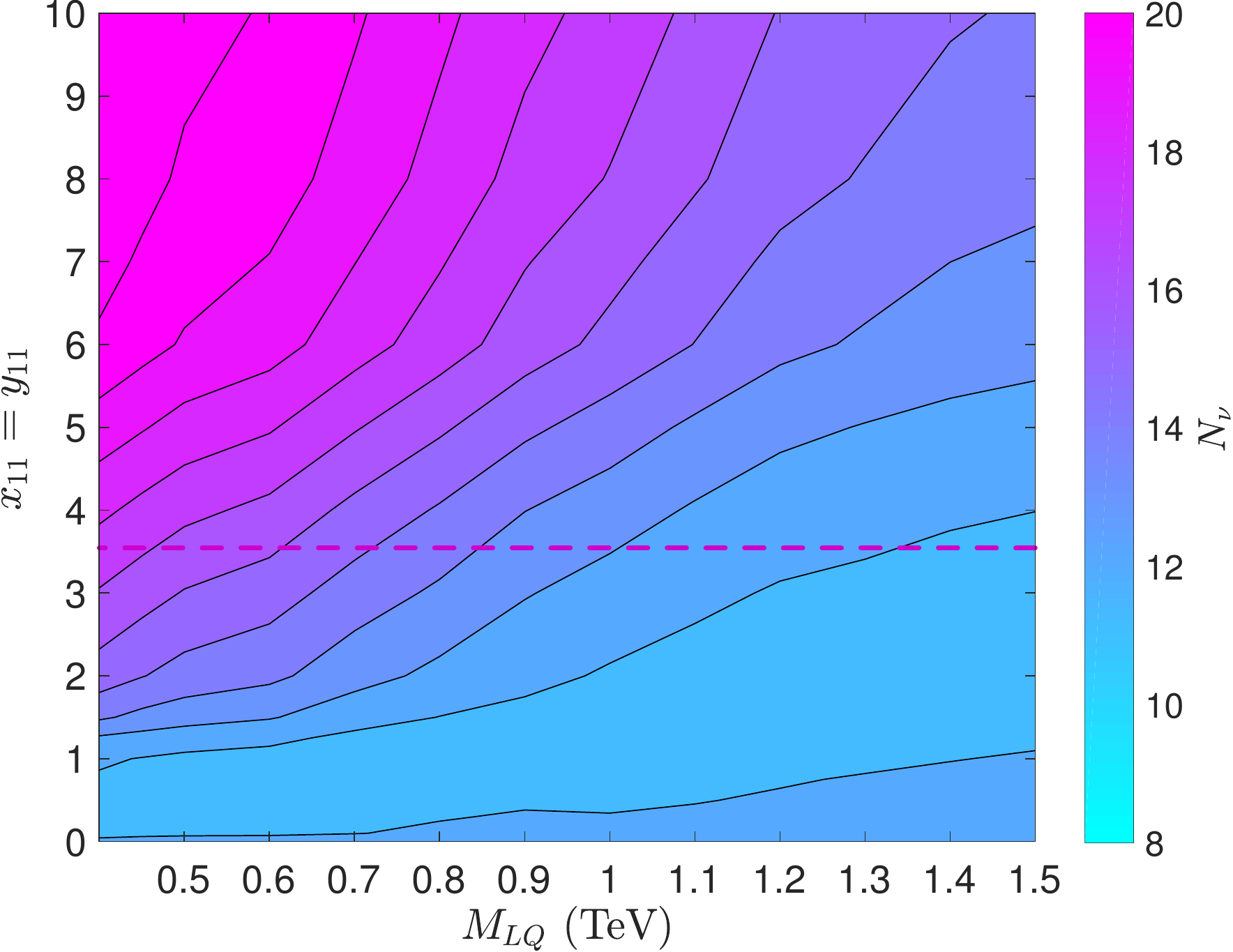} 
%\end{tabular}
\caption{Constraints on Model B from the 53 HESE IceCube events. Left: exclusion contours as a function of the LQ mass and coupling; the best fit point is shown as a red diamond. Middle: astrophysical neutrino flux necessary to accommodate the additional force mediator. Right: best-fit number of atmospheric background neutrinos in the as a function of the LQ mass and coupling. We do not show the corresponding figure for Model A, since every point is within 1$\sigma$ of the Standard Model. In all the three plots the purple (dashed) horizontal lines represent the perturbativity bound of the corresponding couplings.}
\label{fig:degen}
\end{figure}

\begin{figure}[t]
\includegraphics[scale=0.42]{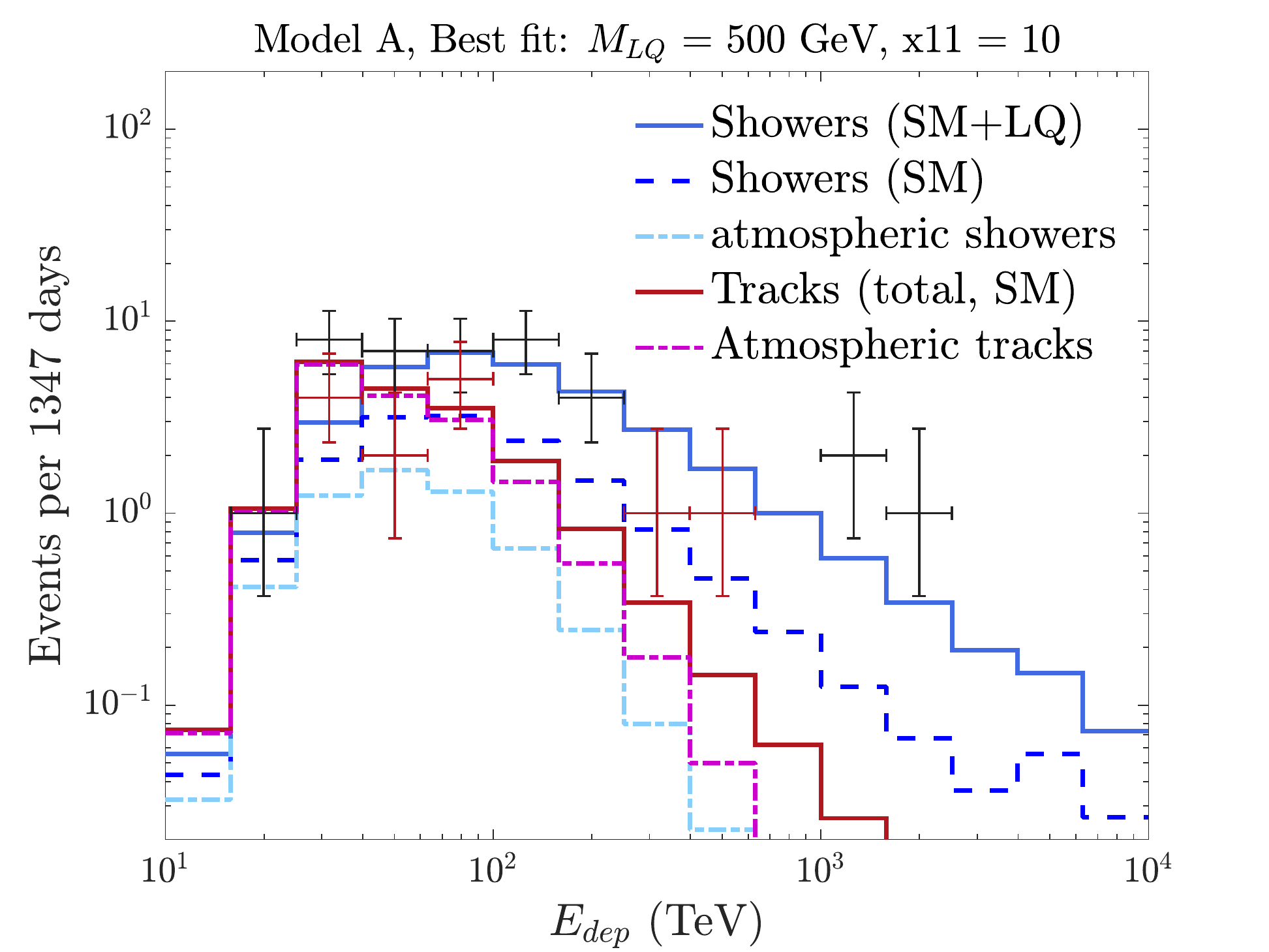} ~
\includegraphics[scale=0.42]{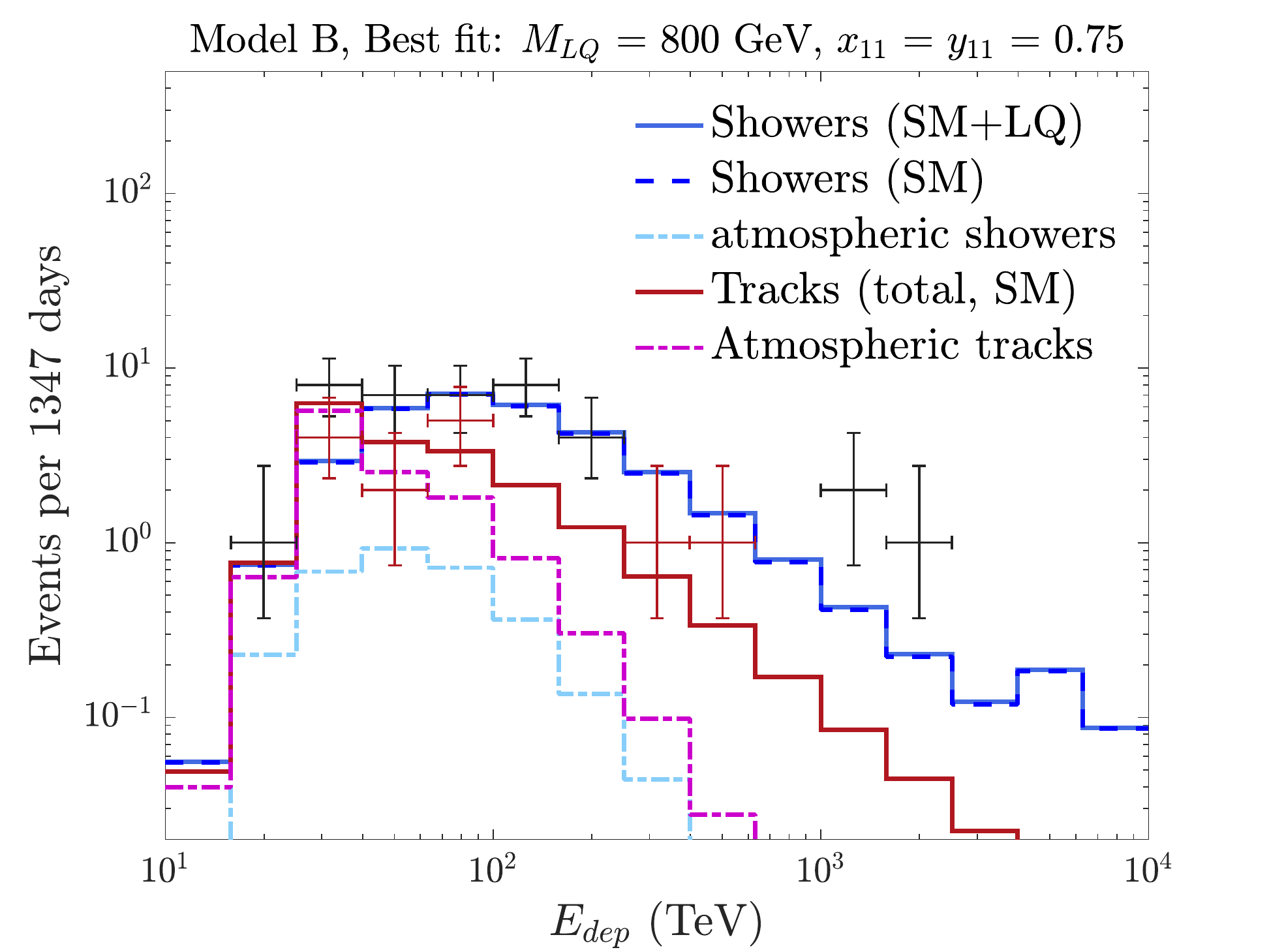}
\caption{Spectrum as expected at IceCube for the best-fit point in Model A (left) and Model B (right). Atmospheric backgrounds are separated in to tracks (magenta) and showers (light blue). Total tracks (atmospheric + astrophysical) are shown in red, while the showers (blue) are split into LQ contribution (dash-dotted), SM contribution (dashed) and total (solid). Data points from four years of IceCube data (crosses) are split into tracks (red) and showers (black). Note (1) the bin at $E_{dep} = 300$ TeV contains both a shower and a track event; and (2) although we have not shown them for clarity, the zero-event bins up to $E_{dep} = 10$ PeV are included in our analysis.} 
\label{fig:ICspectra}
\end{figure}

We then perform a scan over LQ masses between 400 and 1500 GeV, and couplings between zero and 10. From these likelihood evaluations, we note the following results:
\begin{enumerate}
\item The addition of a leptoquark-mediated interaction can generally be compensated by a change in the spectral index and overall normalization of the atmospheric flux, with no significant improvement in the $p$-value with respect to the standard model. For Model A, we find no parameter combination with $x_{11} < 10$ that produces a significant change in the fit to the IceCube data. In contrast, Model B produces changes that are large enough to produce a mild exclusion line. It is not possible to obtain an event distribution that reproduces the PeV events without also increasing the expected event rate in the bins where no events were observed. This is mainly because the large required couplings also yield a decay width that is too large, both suppressing and widening the signal. These exclusions are shown in the first panel of Fig. \ref{fig:degen}. It can be seen that they do not reach the 2$\sigma$ ($1-p = 0.95$) threshold.

\item The best-fit astrophysical spectral index varies around the Standard Model value, between $\gamma = 2.7$ and 3, while atmospheric flux is decreased by a factor between 6 and 8, at the point of largest LQ contribution ($M_{LQ} = 400, x = 10$). An additional atmospheric neutrino component is usually required to compensate for the low-energy showers removed by this rescaling. The number of atmospheric muons is unsurprisingly stable, since these overwhelmingly produce track-like events. These degeneracies are shown in the right-hand panels of Fig. \ref{fig:degen}.

\item The addition of a scalar LQ coupling in both models can lead to a slight improvement in the fit to the data, though this is not statistically distinguishable from the zero-coupling (Standard Model) line. In either model, we see improvement in the $p$-value by 0.05, which is not distinguishable from the SM value ($p \sim 0.35$). The exact mass of the best-fit point is also difficult to pinpoint, given the flatness of the likelihood at couplings below $x_{11} \sim$ a few. For Model B, the best fit point is at $M_{LQ} = 800$ GeV, $x_{11}=y_{11} = 0.75$. The resulting spectrum (dashed blue) is shown in the right-hand panel of Fig. \ref{fig:ICspectra}; this is clearly indistinguishable from the SM-only case (solid blue). We have separated out showers (blue, with black data points) from tracks (red). 

\item Model A gives rise to a qualitatively very different solution, though it remains statistically indistinguishable from the SM case: by suppressing the astrophysical flux to $\sim$ 1/3 of its SM value (i.e. $\phi_0 \rightarrow 2.4 \times 10^{-18}$ GeV$^{-1}$ cm$^{-2}$ s$^{-1}$ sr$^{-1}$) , a very strongly-coupled leptoquark ($M_{LQ} = 500$ GeV, $x_{11} = 10$) can yield an event distribution for cascade events that is compatible with observations, while suppressing the expected rate around the Glashow peak. This improvement in the fit is somewhat compensated by a worse fit to the observed track-like events. This leads to an overall small ($\ll 1\sigma$) improvement in goodness of fit. We must furthermore discount this solution, since it is incompatible with the collider constraints we will find in the next section.

\item One may ask whether the addition of a shower-producing LQ coupling can mitigate the need for a large track misidentification rate $p_{mID}$. We find that this is not the case: taking lower values of $p_{mID}$ leads to an overall worse fit, which the LQ contribution is unable to compensate in either model.  

\item Finally, if the spectral index is fixed to $\gamma = 2.5$, near the 4-year best-fit reported by IceCube, such a best fit point is not recovered in either model; rather, the largest $p$-value is found with $x_{11}=y_{11} = 0$, and the overall goodness of fit is reduced in the entire parameter space. In contrast, adopting a two-component power law leads to an improvement in the overall fit, but the best LQ contribution remains at zero.
\end{enumerate}

We end this section by noting that we have not used angular information, nor have we allowed the flavor composition of the astrophysical flux to vary. Ref. \cite{Dev:2016uxj} has shown that this can help, notably by removing the $\bar \nu_e$ component leading to the Glashow peak. We anticipate that angular information would lead to a slight improvement in the significance of our results, whereas allowing the flavor composition to vary would weaken it, and require extra motivation to explain the lack of a $\bar \nu_e$ flux.

%%%%%%%%%%%%%%%%%%%%%%%%%%%%%%%%%%%%%%%%%%%%
\section{Constraints from LHC } \label{sec:LHC}
%%%%%%%%%%%%%%%%%%%%%%%%%%%%%%%%%%%%%%%%%%%%
\begin{figure}[t]
\begin{center}
\includegraphics[scale=0.6]{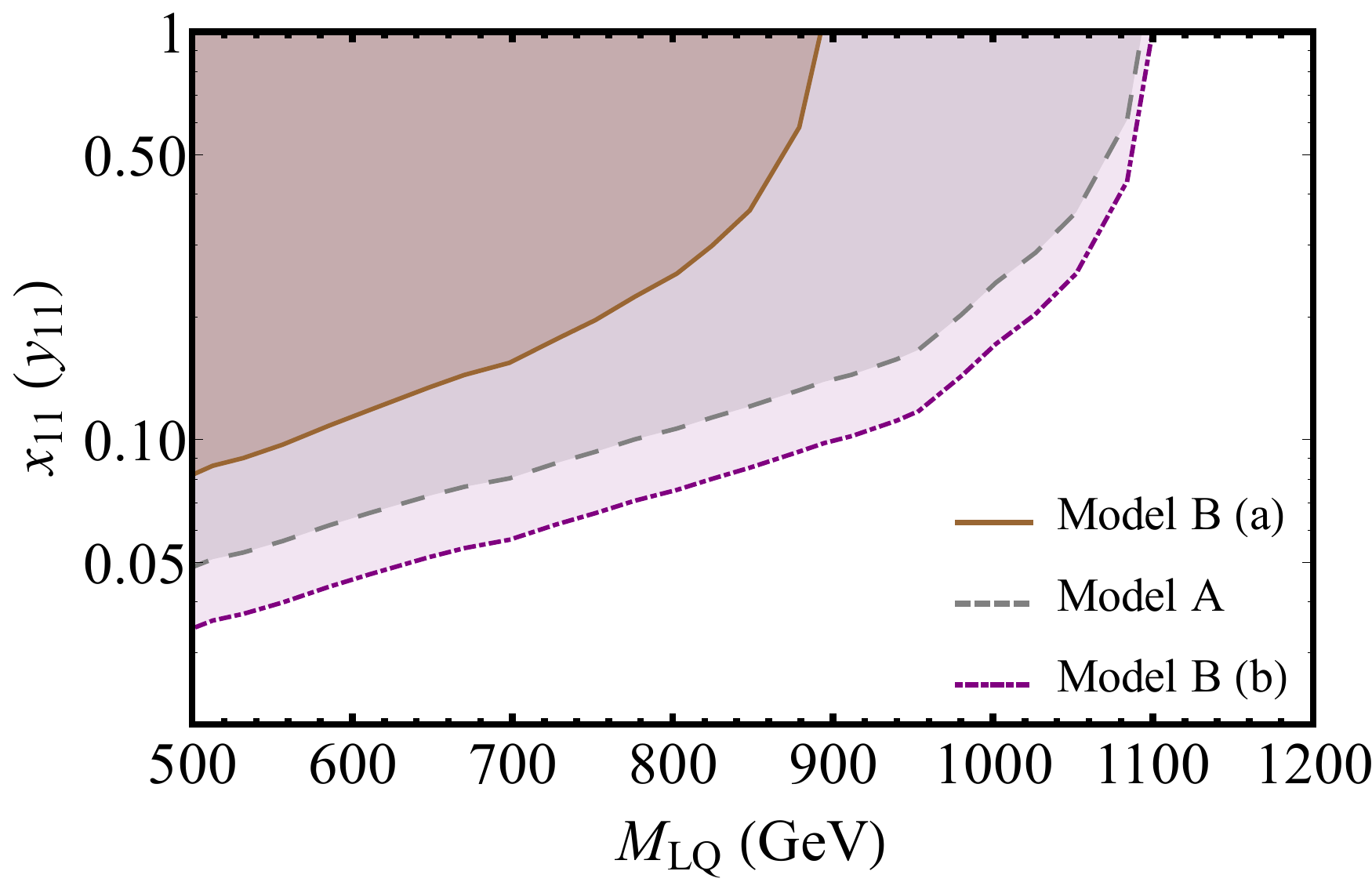}
\caption{Constraints on the relevant couplings $x_{11}, y_{11}$ from the CMS search for the first generation of leptoquark. Shaded region corresponds to the excluded region at the $95\%$ C.L \cite{Aaboud:2016qeg}. Gray dashed line corresponds to the exclusion limit for Model A and the solid brown and purple lines represent Model B. {For Model B, the two lines correspond to the two choices of the parameters (a), (b) defined in the text.}  }
\label{fig:eejj}
\end{center}
\end{figure}

The leptoquarks corresponding to both models can be produced at the LHC and can be detected via its distinct signatures~\cite{Kohda:2012sr, Dorsner:2014axa, Queiroz:2014pra, Allanach:2015ria, Mandal:2015vfa, Mandal:2015lca, Dorsner:2016wpm, Faroughy:2016osc, Raj:2016aky, Dumont:2016xpj, Greljo:2017vvb, Das:2017kkm}. Dedicated searches for the first two generations of leptoquarks
that couple with electrons or muons have been performed at the LHC \cite{Aaboud:2016qeg}. CMS has constrained the leptoquark mass $M_{\rm{LQ}} \geq 1100$ GeV assuming a $100 \%$ branching ratio of leptoquark decaying into a charged lepton and jet. Previous $\sqrt{s} = 8$ TeV searches for first generation LQs have also looked for channels containing two electrons and at least two jets or an electron, a neutrino, and at least two jets~\cite{CMS:2014qpa}. Recently, searches for scalar leptoquark coupled to $\tau$ and $b$-jets have resulted in the constraint $M_{LQ} \geq 850$ GeV (again assuming a 100$\%$ branching ratio). Additionally, in the models considered, the leptoquark can also decay into a neutrino and a quark, giving rise to {multijet} final states associated with large missing energy. Therefore a number of SUSY searches for multijet and MET events can be used to further constrain the model's parameters.

\subsection{Di-lepton + Dijet} 

 In Model A, the leptoquark $\chi^{2/3}_1$ can be pair-produced and decays to a charged lepton and quark. A similar signal can be mimicked by $\chi^{2/3}_1$ and $\chi^{5/3}_2$ in Model B. We consider the pair production of leptoquarks and the following signal topologies 
\begin{eqnarray}
p p &\to & \chi^{2/3}_1 \chi^{-2/3}_1 \to l_i q \bar{l}_j \bar{q'}~~  (\rm{Model\, A}), \\
p p &\to & \chi^{5/3}_2 \chi^{-5/3}_2 / \chi^{2/3}_1 \chi^{-2/3}_1 \to l_i q \bar{l}_j \bar{q'} ~ ~ ({\rm{Model \,B}}).
\label{eejj}
\end{eqnarray}
{In the above,  $l=e,\mu$ and we consider the light quarks i.e., $q,q'=u,d$ and $c$, which lead to the most conservative limits on the couplings}. Below, we derive the limits on the relevant couplings $x_{ij}/y_{ij}$ using the combined 8 TeV and 13 TeV limit on the branching ratios of a leptoquark decaying into a charged lepton and a quark \cite{Aaboud:2016qeg}. The branching ratio of $\chi^{2/3}_1$ for Model A is 
\begin{equation}
\text{Br}( \chi^{2/3}_1 \to l_j q_i) = \frac{M_\chi}{16 \pi}  \frac{x^2_{ij}}{\Gamma({\chi^{2/3}_1})},
\label{brlj}
\end{equation}
where $\Gamma({\chi^{2/3}_1})$ is the total decay width of $\chi^{2/3}_1$ and has the form
\begin{eqnarray}
\Gamma({\chi^{2/3}_1})=\frac{M_\chi}{16 \pi} (x^2_{11}+x^2_{12}+x^2_{13}).
\label{widthchi23}
\end{eqnarray}
%For simplicity, we assume $x_{ij}=0$, for $i>1$, i.e., {\textcolor{red}{the leptoquark only interacts with the first generation of quarks.}} 
With the experimental limit on the branching ratio of LQ$\to e j$ denoted as $\beta$, and using the observed limit from \cite{Aaboud:2016qeg}, an upper limit on $x_{11}$ can be derived as
\begin{eqnarray}
x^2_{11} \le \frac{\beta}{1-\beta} (x^2_{12}+x^2_{13}).
\label{x11limitmodelA}
\end{eqnarray}
These are the same couplings $x_{1i}$ that also contribute to the IceCube events. 

For Model B, the leptoquark $\chi^{2/3}_2$ can decay to both $l j$ and $\nu j$ final states, {hence it can potentially be constrained both from IceCube and from the LHC.} The constraints on the coupling $y_{11}$ that connects $\chi_{2}^{2/3}$ with electron $e$ and light quark result in 
\begin{eqnarray}
y^2_{11} \le \frac{\beta}{1-\beta} \left(\sum_{j=1,2,3} x^{2}_{1j}+\sum_{j=1,2,3} x^{2}_{2j}\right), 
\label{x11limitmodelB}
\end{eqnarray}
where we have considered $\chi_{2}^{2/3}$ interacts only with $e$ and $d$ quark; and with light neutrinos (all flavors) and $u,c$ quarks. Similar expressions can also be derived for muons.

We show the limits on the couplings in Fig.~\ref{fig:eejj}, for both models. The lines correspond to the limits on the branching ratios of the leptoquarks decaying into an electron and jet pair \cite{Aaboud:2016qeg}. Comparable limits can be derived for the muon final state.

\begin{figure}[t]
\begin{center}
\includegraphics[scale=0.5]{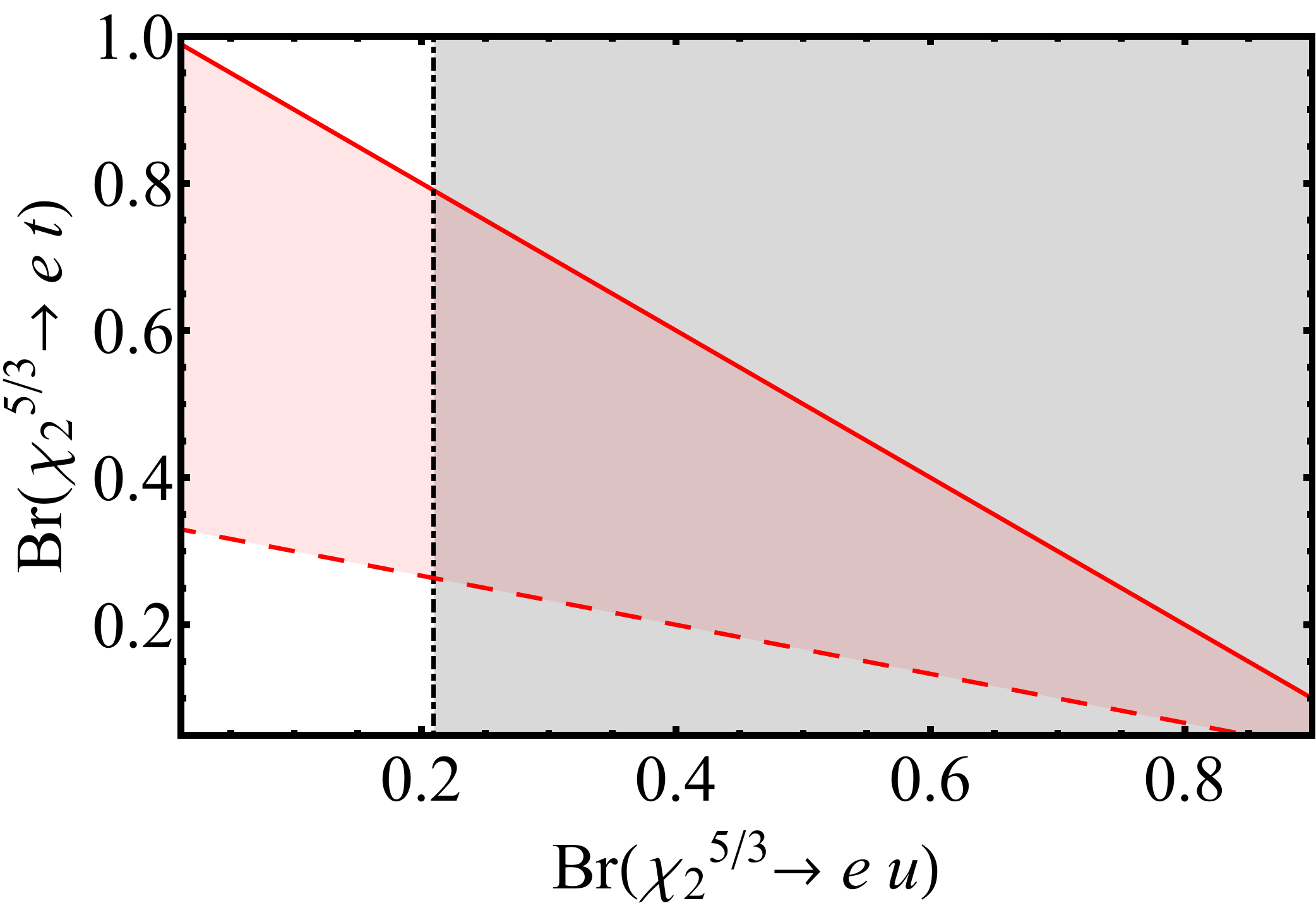}
\caption{Correlation between the two branching ratios of $\chi^{5/3}_2 \to t u$ and $\chi^{5/3}_2 \to e u$. Gray shaded region corresponds to the excluded region from CMS $eejj$ search for first generation of leptoquark, for $M_{LQ} = 650$ GeV. Similar limits hold for leptoquark connecting $\mu$. Solid line corresponds to the scenario of $x_{31}=0.1, x_{11} \neq 0$, while the dashed line correspond to $x_{31}=x_{32}=x_{33}=0.01, x_{11} \neq 0$. For both the lines we consider $x_{11}$ variation between 0.001-1.  }
\label{fig:cor1B}
\end{center}
\end{figure}

\begin{itemize}
\item
For Model A we assume $x_{12}=x_{13}=0.1$, while we vary the leptoquark mass $M_{\chi^{1/3}_1}$ and the coupling $x_{11}$. The gray region is excluded from the letptoquark search in the $eejj$ channel\footnote{{In Figs.~\ref{fig:eejj} and \ref{f:dijetmet}, we represent the leptoquark mass as $M_{LQ}$, i.e., for Model A, and Model B, this represents $M_{\chi^{1/3}_1}$ and $M_{\chi^{2/3}_2}$, respectively.}}. In this scenario, IceCube cannot put stringent constraint on the relevant coupling $x_{11}$ and the mass of  ${\chi^{1/3}_1}$. The most stringent bound appears from the search presented in \cite{Aaboud:2016qeg}. 

\item
For Model B we consider the following illustrative benchmarks
 (a) both couplings $x_{11}$ and $y_{11}$ are equal, i.e. we set $x_{11}=y_{11}$ and  $x_{12}=x_{13}=x_{21}=x_{22}=x_{23}=0.1$ and,  (b)  $x_{11}, y_{11} \neq 0$,  $x_{11}=0.1$, while all other couplings are zero. We vary the leptoquark mass $M_{\chi^{2/3}_2}$ and the relevant coupling $y_{11}$. The disallowed regions correspond to the area covered by the brown solid (for (a)) and purple dotted lines (for (b)) in Fig.~\ref{fig:eejj}~\footnote{For other choices of coupling the bound on $y_{11}$ will be rescaled as ${x}$, hence, naively follow a simple rescaling.}.

{{As discussed in Sec.~\ref{icecube1}, the best-fit data point for Model B is $M_{\chi^{2/3}_2}=800$ GeV with $x_{11}=y_{11}=0.75$. Assuming all other couplings to be zero leads to $\mathrm{BR}(\chi^{2/3} \to e j) \simeq 47\%$. However, this parameter point is already ruled out by existing searches at the LHC \cite{Aaboud:2016qeg}. Even for other leptoquark masses, and with $x_{11}=y_{11}$, the branching ratio of leptoquark decaying to electron and jet remains similar. Following \cite{Aaboud:2016qeg}, this imposes a limit $M_{\chi^{2/3}} \gsim 882$ GeV. A leptoquark with relatively higher mass $M_{\chi^{2/3}_2} \sim $ TeV is still unconstrained from the  present LHC searches, while being in agreement with the 4-year HESE data from IceCube.}}

\end{itemize}

Additionally, Model B can also receive further constraint from the  leptoquark $\chi^{5/3}_2$. The bound on the mass of $\chi^{5/3}_2$ is even more stringent $M_{\chi^{5/3}_2} \geq 1.1$ TeV \cite{Aaboud:2016qeg} (as Br $\chi^{5/3}_2 \to e j=100\%$) for  $x_{11}=y_{11}$ and all other couplings to be zero. The severe bound can  however be relaxed if additional interactions with the  third-generation quarks  i.e., $x_{31}\bar{t}\chi^{5/3}_2e$ or $x_{32}\bar{t}\chi^{5/3}_2\mu$ are present. For a substantial branching ratio $\chi^{5/3}_2 \to t e/ \mu$, the limit on the branching ratios of $\chi^{5/3}_2 \to u e/\mu$ will be relaxed, resulting in a weaker constraint on the $\chi^{5/3} u e $ coupling. We show this in {Fig.~\ref{fig:cor1B} for the scenario $x_{11}, x_{31}\neq$ 0}. {{The LQ state $\chi^{5/3}_2$ however does not couple to light neutrinos and hence is not relevant for the IceCube analysis. }}

\subsection{Dijet + MET} 

The leptoquarks of Model A and Model B can further be constrained from multijet searches. {We adopt the 13 TeV ATLAS search  \cite{Aaboud:2016zdn} and derive the constraints on the mass and coupling using CheckMATE \cite{Drees:2013wra}.} The leptoquarks $\chi^{1/3}_1$ and $\chi^{2/3}_2$  mediate the processes 
\begin{eqnarray}
p p &\to & \chi^{1/3}_1 \chi^{-1/3}_1 \to q_i \nu \bar{q}_j \bar{\nu}~~  (\mathrm{Model \, A}), \\
p p &\to & \chi^{2/3}_2 \chi^{-2/3}_2  \to \nu q_i \bar{\nu} \bar{q}_j ~ ~ ({\mathrm{Model\,  B}}).
\label{djmet}
\end{eqnarray}

\begin{figure}[t]
\begin{center}
\includegraphics[scale=0.85]{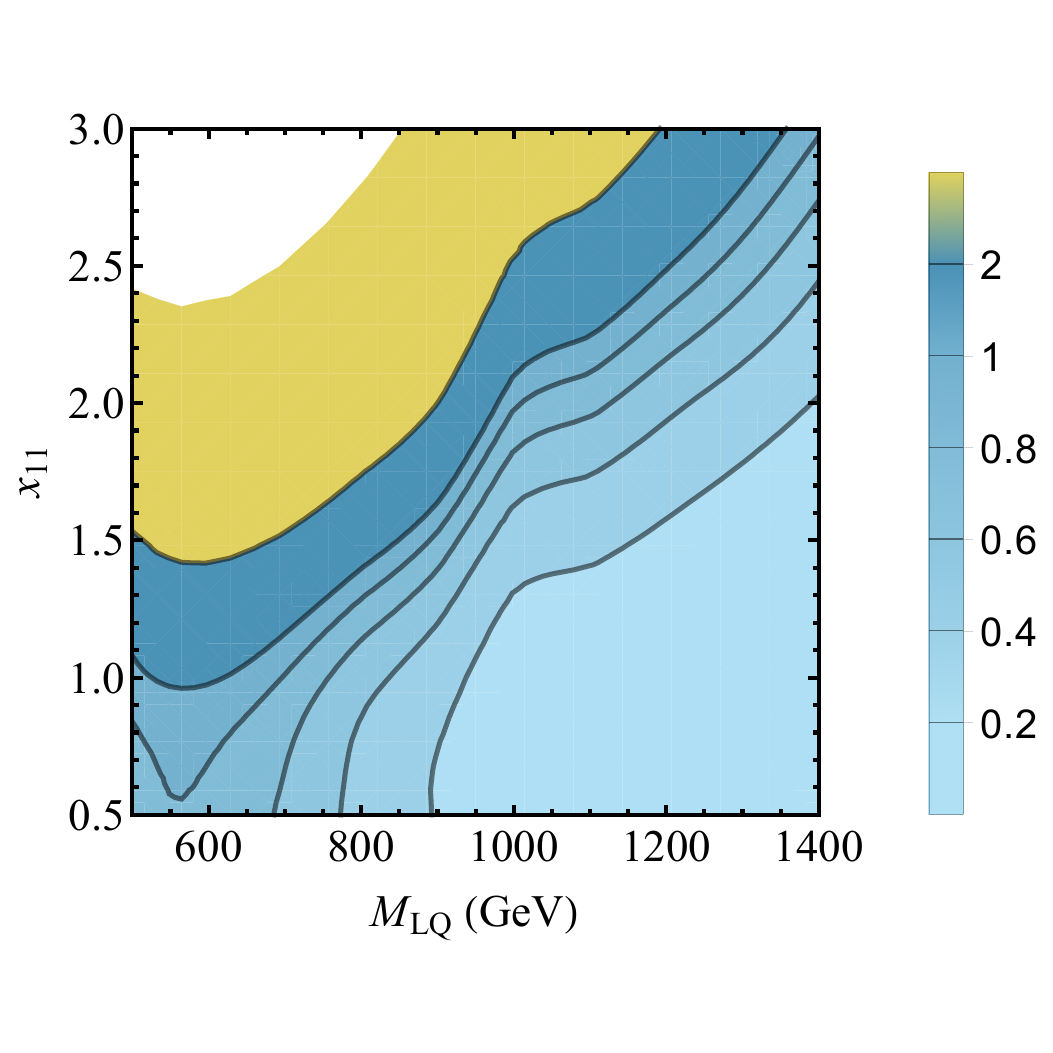}
\caption{The contours for different $r$ (see Eq. (\ref{rvalue})) that correspond to the dijet+MET limits \cite{Aaboud:2016zdn}, relevant for   Model A. The region with $r \geq 1$ is excluded at 95$\%$ C.L.}  %\AV{ $r \geq 1$ is eclcuded (?)}.}
\label{f:dijetmet}
\end{center}
\end{figure}

 For Model A, we consider that the couplings $x_{12}$ and $x_{13}$ are fixed to the benchmark values 0.1. For lower mass and large couplings, strong bounds exist on the leptoquark mass. In Fig.~\ref{f:dijetmet}, we show the contour plots for the 
statistical parameter $r$, defined as 
\begin{eqnarray}
r=\frac{(S-1.96 ~\Delta S)}{S^{95}_{exp}}\,,
\label{rvalue}
\end{eqnarray}
where $S$ and $\Delta S$ represent the signal events and their uncertainty, and the numerator denotes the $95\%$ C.L limit on the number of signal events from {CheckMATE} and the denominator $S^{95}_{exp}$ determines the experimental limit. The excluded regions correspond to $r \ge 1$. A number of cuts have been used to derive the limits, such as, the jet transverse momentum $p_T(j)$, the missing transverse energy $E^{miss}_T$, $m_{eff}$ and the $\Delta \Phi ( jet, E^{miss}_T)$. See \cite{Aaboud:2016zdn} for further detail. 

For lower masses such as $M_{LQ}=600$ GeV, the coupling $x_{11} \ge \mathcal{O}(1)$ is  ruled out, as shown in Fig.~\ref{f:dijetmet}. {With increasing leptoquark mass, such as, TeV or larger, the limit on $x_{11}$ is strongly relaxed.} %For larger masses $M_{LQ}=1200$ GeV, much larger couplings are still allowed. 
{A comparison between di-lepton+dijet (Fig.~\ref{fig:eejj}) and jet+MET (Fig.~\ref{f:dijetmet}) for Model A shows that  $eejj$ gives a stronger bound on the leptoquark coupling $x_{11}$.} For Model B, we do not  explicitly show  the constraints on the parameter space. For the best-fit points  $x_{11}=y_{11}=0.75$ and $M_{LQ}=800$ GeV as discussed in Sec.~\ref{sec:ICresults}, the branching ratio $\chi^{2/3}_2 \to j \nu$ is predicted to be $49.3\%$, which leads to coupling $y_{11}$ being largely unconstrained.

In addition to the collider constraints, leptoquark models 
can also be constrained from different flavor violating observables, such as, $\mu \to e \gamma$, $\mu-e$ conversion in nuclei etc. This has been extensively studied in the literature \cite{Gabrielli:2000te, Arnold:2013cva}.  Typically, the tightest constraint arises from $\mu-e$ conversion in nuclei that bounds $|x_{11}x_{12}| \lsim 10^{-5}$ for a  $\sim$ TeV LQ in Model A. However, Model B with the choice $x_{11} \neq 0$ and all other couplings zero as has been considered in Sec.~\ref{icecube1}  will remain unaffected from these constraints.

%%%%%%%%%%%%%%%%%%%%%%%%%%%%%%%%%%%%%%%%%%%%%%%
\section{Flavor anomalies}  
\label{sec:flavor}
%%%%%%%%%%%%%%%%%%%%%%%%%%%%%%%%%%%%%%%%%%%%%%%
Recent LHCb observations on rare $B$-meson decays show $\sim 2.5 \sigma$ deviations from the SM in the observables $R_{K}$ and $R_{K^{\ast}}$~\cite{Aaij:2014ora, Aaij:2017vbb}. There are quite a few studies~\cite{Hiller:2014yaa, Gripaios:2014tna, Varzielas:2015iva, Becirevic:2015asa, Fajfer:2015ycq, Becirevic:2016yqi, Aaij:2015oid, Wehle:2016yoi, Becirevic:2016oho, Das:2016vkr, Crivellin:2017zlb, Becirevic:2017jtw, Hiller:2017bzc, Cai:2017wry, Aloni:2017ixa,DiLuzio:2017vat} which explain these deviations using the leptoquarks of Model B. We therefore deem it relevant to add a short discussion on the implications of these flavour anomalies on the leptoquark models that we are considering.

Leptoquarks with hypercharge $Y = \mathbf{7/6}$ from Model B, have been proposed to explain the flavor anomalies in the 
$B \to K l l$ decays \cite{Becirevic:2017jtw}, where the transition occurs through one-loop processes, mediated by the leptoquark $\chi^{5/3}_2$.  
This can successfully explain both the $R_K$ and $R^{*}_K$ anomalies. In order to explain the experimental results, fairly large couplings $x_{22}$, %c\mu}, 
$x_{32}$ %t\mu}$ 
are  required, close to the edge of the perturbative regime. Below, we examine in detail the collider constraints 
on such large coupling. The partial decay width of the leptoquark decaying in the  $c l$ and $t l$ states have  the following form:

\begin{align}
\Gamma (\chi^{5/3}_2 \to l t) = \frac{|x_{3l}|^2}{8 \pi M^3_{\chi^{5/3}_2}}\left(M^2_{\chi^{5/3}_2}-m^2_t\right)^2, ~~~~
\Gamma (\chi^{5/3}_2 \to l c) = \frac{|x_{2l}|^2}{8 \pi M^3_{\chi^{5/3}_2}}\left(M^2_{\chi^{5/3}_2}-m^2_c\right)^2.
\end{align}
%\end{subequations}
%
We consider the following two scenarios: 
\begin{itemize}
\item Only $x_{22}$ and $x_{32}$ are non-zero and fairly large $|x_{22}|, |x_{32}| \ge 1.0$.  This is required to satisfy the flavor anomalies and is in agreement with \cite{Becirevic:2017jtw}. As an illustrative example, we consider two masses $m_{\chi^{5/3}_2}=650$ and $1000$ GeV.

\item Additionally, the couplings including top-quark and $\tau$ is non-zero and large, i.e., $x_{33}>  x_{32},  x_{22} $, while $|x_{32}| \sim |x_{22}|$. 
\end{itemize}
We show the branching ratios of $\chi^{5/3}_2$ decaying into $c \mu$, $t \mu$ and $t \tau$ states in Table.~\ref{tablebench}, and further compare the benchmark choices with the di-lepton+dijet search \cite{Aaboud:2016qeg}. 

\begin{table}[h]
\centering
\begin{tabular}{|c|c|c|c|c|c|c|}
\hline
\begin{tabular}[c]{@{}c@{}}$M_{\chi_{2}^{5/3}}$ \\ (GeV)\end{tabular} & $x_{22}$ & $x_{32}$ & $x_{33}$ & Br($\chi_{2}\to c\mu$)  & Br($\chi_{2}\to t\tau$) & Br($\chi_{2}\to t\mu$) \\ \hline \hline
650 (*)                                                               & $-2.4$        & $1.0$   & 0        & 0.87 (\textgreater Br1) & 0                       & 0.13                   \\ \hline
650                                                                   & 2        & $-1.5$   & 5        & 0.15 (\textless Br1)    & 0.78                    & 0.07                   \\ \hline
1000 (*)                                                                  & $-3.4$        & 1.2   & 0        & 0.89 (\textgreater Br2)    & 0                       & 0.10                   \\ \hline
1000                                                                  & 3.4        & $-1.4$   & 5        & 0.31 (\textless Br2)    & 0.64                    & 0.05                   \\ \hline
\end{tabular}
\caption{Predicted branching ratios for benchmark values of the mass of the leptoquark $\chi^{5/3}_2$ and the couplings $x_{22}, x_{32}$ and $x_{33}$. Note that the scenario $*$ is ruled out by the leptoquark search for $p p \to LQ ~ LQ  \to \mu \mu j j$~\cite{Aaboud:2016qeg}. The experimental limit on the branching ratios Br1 and Br2 are 0.25 and 0.86, respectively, following \cite{Aaboud:2016qeg}.}
\label{tablebench}
\end{table}

 These above benchmark values of the couplings ensure the $1.5\sigma$ agreement in the measurement of $R_{K}$ and $R_{K^{\ast}}$~\cite{Becirevic:2017jtw}. However, these can be further constrained  from other LHC searches, in particular di-lepton+dijet. In the absence of coupling $x_{33}$, the  lower leptoquark mass 650 GeV with large $x_{32}$ and  $x_{22}$ gives a sizeable branching ratio Br$(\chi^{5/3}_2 \to c \mu)=87\%$, larger than the experimental limit $\rm{Br1} \le 25\%$. For heavier or lower masses with additional $\chi^{5/3}_2 \to t \tau$ decay mode open, the branching ratio into $\chi^{5/3}_2 \to c \mu$ is yet unconstrained. Hence, a heavier leptoquark or lighter leptoquark with additional coupling with third-generation quark and lepton can successfully explain both the $R_K-R^*_K$ anomalies, while remaining consistent with di-lepton+dijet search. The leptoquark $\chi^{5/3}_2$ does not couple with light neutrinos. Hence, a direct comparison with the IceCube constraint can not be performed.
In addition to the above processes, leptoquarks can further be constrained from atomic parity violation (APV), $K_L \to \mu^- e$ and $D^0 \to \mu^- e$ rare decays \cite{Dorsner:2014axa, Raj:2016aky, Mileo:2016zeo}. The constraint from $D^0$ decays on the product of Yukawa is rather weak. The $K_L \to \mu e$ branching ratio is bounded as $4.7 \times 10^{-12}$. This puts stringent constraint on $\chi^{2/3}_{2}$ as $|y_{22} y^*_{11}| \le 2.1 \times 10^{-5}$ for a TeV scale $\chi^{2/3}_2$ \cite{Dorsner:2014axa} from Model B (similar constraint holds for $\chi^{2/3}_1$ for $\chi_{1}^{2/3}$ of Model A). Additionally bounds from APV translates as $|y_{11}| \le 0.3( M_{LQ}/1 \rm{TeV})$. This is more stringent in the higher mass region $M_{LQ} \sim $ TeV. A comparison between Fig.~\ref{fig:eejj} and the APV bound \cite{Raj:2016aky} shows that for lower mass, the 13 TeV LHC limit from the $l l j j$ search \cite{Aaboud:2016qeg} puts more stringent limit. For detailed discussion on APV bound, see \cite{Dorsner:2014axa, Raj:2016aky}.

%%%%%%%%%%%%%%%%%%%%%%%%%%%%%%%%%%%%%%%%%%%%%%%%%%%%%
\section{Conclusion} \label{sec:conclusion}
%%%%%%%%%%%%%%%%%%%%%%%%%%%%%%%%%%%%%%%%%%%%%%%%%%%%%
In this work, we have analyzed the constraints on scalar leptoquarks with hypercharge $Y = \mathbf{1/6}$ and $\mathbf{7/6}$  that arise from neutrino as well as collider experiments. Strongly interacting scalar leptoquarks can contribute 
to the high-energy IceCube events, which we have explored in detail. 

Previous studies have shown that such a new scalar can successfully improve fits to the high-energy PeV events.  However, we show that this generically tends to worsen the fit to bins where zero events were observed, once  the entire range of 10 TeV to 10 PeV is taken into account. Furthermore, we have shown that self-consistently allowing the astrophysical and atmospheric flux to vary strongly reduces the power to distinguish a  leptoquark signal. Once the full range  is taken into account,  the high-energy events seen at IceCube  do not carry sufficient statistical power to exclusively confirm or exclude the presence of the scalar leptoquark. 

We  further consider a number of LHC searches, such as collider constraint for first-generation leptoquark in the $eejj$ final state, the dijet+met constraint, as well as the recently reported flavor anomalies $R_K$ and $R_{K^*}$ by LHCb. We show that both light and heavier leptoquarks in the 400-1200 GeV mass range can be in agreement with the IceCube events. However, lighter leptoquark states, e.g. with masses around 500 GeV, are further constrained from the $eejj$ search. Heavier leptoquarks in the TeV mass range, can be in agreement with both  IceCube and  LHC. We further examine the models in the context of the $B \to K l l$ anomalies and find that the low mass leptoquark can consistently explain the flavor anomalies and can be in agreement with di-lepton+dijet bound, if the branching ratio of the leptoquark decaying to a charm and muon is less that 10$\%$.

%%%%%%%%%%%%%%%%%%%%%%%%%%%%%%%%%%%%%%%%%%%%%%%%%%%%%%%%%%%%%%%
%{\bf{Acknowledgement:}} 
\noindent
{\bf Acknowledgements\,:}\\
U. K. D. acknowledges the support from the Department of Science and Technology, Government of India under the fellowship reference number PDF/2016/001087 (SERB National Post-Doctoral Fellowship).  M. M. acknowledges the support from DST INSPIRE Faculty Research grant (IFA14-PH-99) of India  and Royal Society International Exchange Programme. M. M. thanks IPPP, Durham University, UK for their hospitality. D. K. thanks National Research Foundation, South Africa for support in forms of CSUR and Incentive funding. A. C. V. is supported by an Imperial College Junior Research Fellowship.
%%%%%%%%%%%%%%%%%%%%%%%%%%%%%%%%%%%%%%%%%%%%%%%%%%%%%%%%%%%%%%%

\appendix

\section*{Appendix}
Here we present all the relevant expressions for the neutrino-nucleon cross section for both Model A and Model B.

\section{Neutrino-nucleon interactions for Model A} \label{app:modelA}

For Model A, the relevant interactions for IceCube are the neutrino-nucleon neutral current interactions mediated by the leptoquark. Note that, in this model, the leptoquark $\chi_1$ does not generate any charged current type interactions that can contribute in IceCube. Below, we write all the possible interaction terms, taking into account the SM contribution. 

%%%%\subsection{$\bar{\nu}-N$ interaction}
Considering the NC-type processes (${\nu}_{i}{d}_{j}\to {\nu}_{k}{d}_{l}$ and ${\nu}_{i}\bar d_{j}\to {\nu}_{k} \bar d_{l}$), the differential cross section for neutrino-nucleon\footnote{Note that here $N$ corresponds to either the proton or the neutron; we have not followed the common approach of averaging over nucleons (which yields terms proportional to $(u + d)/2$), since ice is not isoscalar. } ($\bar{\nu}$-$N$) interaction will be given by, 
\begin{align}
\frac{d^2\sigma_{\nu_j}}{dxdy}= \frac{m_N E_{\nu}}{16 \pi} x \sum_{i}\Bigg[&\left\lbrace |a_{f_{ij}}|^2+|b_{f_{ij}}|^2 (1-y)^2 \right\rbrace q_i(x,Q) + \left\lbrace |{b_{f_{\bar{ij}}}}|^2+|a_{f_{\bar{ij}}}|^2 (1-y)^2 \right\rbrace \bar q_i(x,Q) \Bigg],
\label{eq:dsignubar}
\end{align}
where the sum is over the quark flavors, and the PDFs are: $q_i = u, d, c, s, b$ (we ignore top contributions) and  $ \bar q_i = \bar u, \bar d, \bar c, \bar s, \bar b$. The $a_{f_{ij}}$ and $b_{f_{ij}}$ coefficients contain the information about the interactions (both SM and LQ) between neutrino flavor $j$ and quark flavor $i$. For $\chi_{1}^{1/3}$ mediated processes only down-type quarks take part. Dropping the neutrino ($j$) index when the LQ interaction is not present, and recalling $G_F \equiv \sqrt{2} g^2/8m_W^2$, these coefficients are:
\begin{subequations}
\begin{align}
a_{f_{u}} &= \frac{g^2}{(1-\sin^2 \theta_w)}\frac{L_{f_u}}{(Q^2(x,y,E_{\nu})+m^2_Z)} = a_{f_{\bar{u}}} \\
b_{f_{u}} &= \frac{g^2}{(1-\sin^2 \theta_w)}\frac{R_{f_u}}{(Q^2(x,y,E_{\nu})+m^2_Z)} = b_{f_{\bar{u}}}\\
a_{f_{d}} &= \frac{g^2}{(1-\sin^2 \theta_w)}\frac{L_{f_d}}{(Q^2(x,y,E_{\nu})+m^2_Z)} = a_{f_{\bar{d}}}\\
b_{f_{dj}} &= \frac{g^2}{(1-\sin^2 \theta_w)}\frac{R_{f_d}}{(Q^2(x,y,E_{\nu})+m^2_Z)} - 
\frac{x_{dj}x_{dk}}{xs(E_{\nu})-M_{LQ}^{2}} \\
b_{f_{\bar{dj}}} &= \frac{g^2}{(1-\sin^2 \theta_w)}\frac{R_{f_d}}{(Q^2(x,y,E_{\nu})+m^2_Z)} - 
\frac{x_{dj}x_{dk}}{Q^2(x,y,E_{\nu})-xs(E_{\nu})-M_{LQ}^{2}}\;.
\end{align}
\end{subequations}
$a_{f_{c}}  = a_{f_{u}}$ and  $a_{f_{s}}  = a_{f_{b}} =  a_{f_{d}}$, and likewise for the $b$ coefficients. Note that only the $k = j$ components interfere with the SM $Z$ exchange.  
We have used the kinematical quantities:
\begin{align*}
s(E_{\nu})=2 m_{N} E_{\nu},~~~ Q(x,y,E_{\nu}) = \sqrt{2 x y m_{N} E_{\nu}}.
\end{align*}
 Finally,
\begin{gather*}
L_{f_u}=\frac{1}{2}-\frac{2}{3} \sin^2 \theta_w, ~~~~ L_{f_d}=-\frac{1}{2}+\frac{1}{3} \sin^2 \theta_w\;, \\
R_{f_u}=-\frac{2}{3} \sin^2 \theta_w, ~~~~ R_{f_d}=\frac{1}{3} \sin^2 \theta_w\;.
\end{gather*}
Here $y=\frac{E'_{\nu}}{E_{\nu}}$ with $E'_{\nu}=(E_{\nu}-E_l)$ being the energy loss of the incoming neutrino.
The $\bar \nu$-$N$ differential cross section can be written following Eq.~(\ref{eq:dsignubar}) by interchanging $q_i \leftrightarrow \bar q_i$. 
\section{Neutrino-nucleon interactions for Model B} \label{app:modelB} 
%%%%%%%%%%%%%%%%%%%%%%%%%%%%%%%%%%%%%%%%%%%%%%%%%%%%%%%%%%%%%%%%%%
The leptoquark $\chi_{2}^{2/3}$ from Model B can contribute in NC type as well as CC type processes, as shown in  Fig.~\ref{f:feyndiag2}. The relevant neutrino-nucleon cross sections are discussed below.

%%%%%%%%%%%%%%%%%%%%%%%%%%%%%%%%
\subsection{NC type processes}
%%%%%%%%%%%%%%%%%%%%%%%%%%%%%%%%
Considering the NC-type processes ($\bar{\nu}_{i}\bar{u}_{j}\to \bar{\nu}_{k}\bar{u}_{l}$ and $\bar{\nu}_{i}u_{j}\to \bar{\nu}_{l}u_{k}$), the differential cross section for antineutrino-nucleon ($\bar{\nu}$-$N$) interaction in this case is the same as for Model A \eqref{eq:dsignubar}.  For $\chi_{2}^{2/3}$ the NC type processes only up type quarks take part. This time, the explicit forms of $a_{f_i}$ and $b_{f_i}$ are given by,
\begin{subequations}
\begin{align}
a_{f_{u}} &= \frac{g^2}{(1-\sin^2 \theta_w)}\frac{L_{f_u}}{(Q^2(x,y,E_{\nu})+m^2_Z)} = a_{f_{\bar{u}}}\\
b_{f_{uj}} &= \frac{g^2}{(1-\sin^2 \theta_w)}\frac{R_{f_u}}{(Q^2(x,y,E_{\nu})+m^2_Z)} - 
\frac{x_{uj}x_{uk}}{Q^2(x,y,E_{\nu})-xs(E_{\nu})-M_{LQ}^{2}}\\
b_{f_{\bar{u}}} &= \frac{g^2}{(1-\sin^2 \theta_w)}\frac{R_{f_u}}{(Q^2(x,y,E_{\nu})+m^2_Z)} - 
\frac{x_{uj}x_{uk}}{xs(E_{\nu})-M_{LQ}^{2}}\\
a_{f_{d}} &= \frac{g^2}{(1-\sin^2 \theta_w)}\frac{L_{f_d}}{(Q^2(x,y,E_{\nu})+m^2_Z)} = a_{f_{\bar{d}}}\\
b_{f_{d}} &= \frac{g^2}{(1-\sin^2 \theta_w)}\frac{R_{f_d}}{(Q^2(x,y,E_{\nu})+m^2_Z)} = b_{f_{\bar{d}}}\;.
\end{align}
\end{subequations}
Once more, the second- and third-generation couplings are the same as $u$ and $d$, except that we set the LQ couplings $x_{ij}$ to zero. The NC type $\bar \nu$-$N$ differential cross section can be written following the interchanges $q_i \leftrightarrow \bar q_i$.

%%%%%%%%%%%%%%%%%%%%%%%%%%%%%%%%
\subsection{CC type processes} \label{app:CC}
%%%%%%%%%%%%%%%%%%%%%%%%%%%%%%%%
Since the LQ $\chi_{2}^{2/3}$ can also lead to charged leptons in the final state of neutrino-quark interactions it will lead to CC type processes (e.g., $\nu_{i}\bar{u}_{j}\to e_{k}\bar{d}_{l}$ and $\nu_{i}d_{j}\to e_{k}u_{l}$) also, see Fig.~\ref{f:feyndiag2}. The differential cross section for the neutrino nucleon cross section in this case is given by,
\begin{equation}
\frac{d^2\sigma_{\nu_j}^{CC}}{dxdy}= \frac{m_N E_{\nu}}{16 \pi} x \sum_i \left( |a_{f_{ji}}|^2 q_i (x,Q) + |b_{f_{j \bar i}}|^2 (1-y)^2\bar q_i(x,Q) \right)\;.
\label{eq:dsignu3}
\end{equation}

The $a$'s and $b$'s are:
\begin{subequations}
\begin{align}
a_{f_{u}} &= \frac{g^{2}}{Q^2(x,y,E_{\nu}) + m^2_W} = b_{f_{d}}
\\
b_{f_{u}} &= \frac{g^2}{Q^2(x,y,E_{\nu}) + m^2_W} - 
\frac{x_{uj} y_{kl}}{xs(E_{\nu})-M_{LQ}^{2}}
\\
a_{f_{d}} &= \frac{g^2}{Q^2(x,y,E_{\nu}) + m^2_W} - 
\frac{x_{uj} y_{kl}}{Q^2(x,y,E_{\nu})-xs(E_{\nu})-M_{LQ}^{2}}.
\end{align}
\end{subequations}
The antineutrino-nucleon differential cross section can again be obtained from Eq.~(\ref{eq:dsignu3}) by interchanging the quark and antiquark PDFs.

%%%%%%%%%%   References  %%%%%%%
\bibliographystyle{JHEP}
\bibliography{lqiclhc_ref.bib}
%%%%%%%%%%%%%%%%%%%%%%%%%%%%%%%%

\end{document}